\documentclass[10pt,conference]{IEEEtran}
\IEEEoverridecommandlockouts
\usepackage{cite}
\usepackage{amsmath,amssymb,amsfonts}
\usepackage{algorithmic}
\usepackage{graphicx}
\usepackage{textcomp}
\usepackage{soul}  % 导入 soul 包
\usepackage{booktabs} 
\usepackage{multirow} 
\usepackage{tabularx}
\usepackage{amsmath}
\usepackage{makecell}
\usepackage{amsmath,amsfonts}
\usepackage{algorithmic}
\usepackage{algorithm}
\usepackage{array}
\usepackage[caption=false,font=normalsize,labelfont=sf,textfont=sf]{subfig}
\usepackage{threeparttable}
\usepackage{textcomp}
\usepackage{stfloats}
\usepackage{url}
\usepackage{verbatim}
\usepackage{graphicx}
\usepackage{cite}
\usepackage{booktabs} 
\usepackage[table,xcdraw]{xcolor}
\usepackage{diagbox}
\usepackage{multirow} 
\usepackage{lipsum} % 用于生成示例文本
\usepackage{booktabs} % 用于漂亮的表格线
\usepackage[utf8]{inputenc}
\usepackage{fontawesome5} % 导入 fontawesome 宏包
\usepackage[many]{tcolorbox} % for summary box
\newtcolorbox{boxK}{
    top=2pt,
    bottom=2pt,
    left=2pt,
    right=2pt,
    boxrule = 0pt,
    toprule = 0pt, % top rule weight
}

\def\BibTeX{{\rm B\kern-.05em{\sc i\kern-.025em b}\kern-.08em
    T\kern-.1667em\lower.7ex\hbox{E}\kern-.125emX}}

\makeatletter
\newcommand{\linebreakand}{%
  \end{@IEEEauthorhalign}
  \hfill\mbox{}\par
  \mbox{}\hfill\begin{@IEEEauthorhalign}
}
\makeatother

\begin{document}

\title{
% On the Robustness of LLM-powered Program Repair: Exploring and Lifting with Metamorphic Testing
Exploring and Lifting the Robustness of LLM-powered Automated Program Repair with Metamorphic Testing
% MT-LAPR: Metamorphic Testing for Automated Program Repair with Large Language Models\\
%{\footnotesize \textsuperscript{*}}
}
\author{Pengyu~Xue\IEEEauthorrefmark{3},
        Linhao~Wu\IEEEauthorrefmark{3},
        Zhen~Yang\IEEEauthorrefmark{4},
        Zhongxing~Yu,
        Zhi~Jin,
        Ge~Li,
        Yan~Xiao,\\
        Shuo~Liu,
        Xinyi~Li,
        Hongyi Lin,
        and Jingwen~Wu% <-this % stops a space

\thanks{{Pengyu Xue, Linhao Wu, Zhen Yang, Zhongxing Yu, Hongyi Lin, and Jingwen Wu are with the School of Computer Science and Technology, Shandong University, Qingdao, China. E-mail: \{zhenyang, zhongxing.yu, jingwenwu\}@sdu.edu.cn, \{xuepengyu, wulinhao, 202435246\}@mail.sdu.edu.cn. Zhi Jin and Ge li are with the Key Laboratory of High Confidence Software Technologies (Peking University), Ministry of Education, and the School of Computer Science, Peking University, Beijing, China. E-mail: \{zhijin, lige\}@pku.edu.cn. Xinyi Li is with the School of Electrical and Electronic Engineering, Nanyang Technological University, Singapore. E-mail: LIXI0104@e.ntu.edu.sg. Yan Xiao is with the School of Cyber Science and Technology, Sun Yat-sen University, Shenzhen, China. E-mail: xiaoy367 @mail.sysu.edu.cn. Shuo Liu is with the Department of Computer Science, City University of Hong Kong, Hong Kong, China. E-mail: sliu273-c@my.cityu.edu.hk.}}% <-this % stops a space
\thanks{\IEEEauthorrefmark{3}~Pengyu Xue and Linhao Wu are co-first authors.}% <-this % stops a space
\thanks{\IEEEauthorrefmark{4}~Zhen Yang is the corresponding author.}
}% <-this % stops a space

\markboth{IEEE TRANSACTIONS ON SOFTWARE ENGINEERING}%
{Shell \MakeLowercase{\textit{et al.}}: A Sample Article Using IEEEtran.cls for IEEE Journals}

% \and
% \and
% \and
% \and
% \and

\maketitle

% \author{Pengyu~Xue\IEEEauthorrefmark{3},
%         Linhao~Wu\IEEEauthorrefmark{3},
%         Zhen~Yang\IEEEauthorrefmark{4},
%         Xinyi~Li,
%         Zhongxing~Yu,
%         Zhi~Jin,
%         Ge~Li,
%         Yan~Xiao,
%         and Jingwen~Wu% <-this % stops a space

% \thanks{{Pengyu Xue, Linhao Wu, Zhen Yang, Zhongxing Yu and Jingwen Wu are with the School of Computer Science and Technology, Shandong University, Qingdao, China. E-mail: \{zhenyang, zhongxing.yu, jingwenwu\}@sdu.edu.cn, \{xuepengyu, wulinhao\}@mail.sdu.edu.cn. Zhi Jin and Ge li are with the Key Laboratory of High Confidence Software Technologies (Peking University), Ministry of Education, and the School of Computer Science, Peking University, Beijing, China. E-mail: {zhijin, lige}@pku.edu.cn. Xinyi Li is with the School of Electrical and Electronic Engineering, Nanyang Technological University, Singapore. E-mail: LIXI0104@e.ntu.edu.sg. Yan Xiao is with the School of Cyber Science and Technology, Sun Yat-sen University, Shenzhen, China. E-mail: xiaoy367 @mail.sysu.edu.cn.}}% <-this % stops a space
% \thanks{\IEEEauthorrefmark{3}~Pengyu Xue and Linhao Wu are co-first authors.}% <-this % stops a space
% \thanks{\IEEEauthorrefmark{4}~Zhen Yang is the corresponding author.}
% }% <-this % stops a space

\begin{abstract}
In recent years, Large language model-powered Automated Program Repair (LAPR) techniques have achieved state-of-the-art bug-fixing performance and have been pervasively applied and studied in both industry and academia. Nonetheless, LLMs were proved to be highly sensitive to input prompts, with slight differences in the expressions of semantically equivalent programs potentially causing repair failures. Therefore, it is crucial to conduct robustness testing on LAPR techniques before their practical deployment. However, related research is scarce. To this end, we propose MT-LAPR, a Metamorphic Testing framework exclusively for LAPR techniques, which summarizes nine widely-recognized Metamorphic Relations (MRs) by developers across three perturbation levels: token, statement, and block. Afterward, our proposed MRs are applied to buggy codes to generate test cases, which are semantically equivalent yet to affect the inference of LAPR. Experiments are carried out on two extensively examined bug-fixing datasets, i.e., Defects4J and QuixBugs, and four bug-fixing abled LLMs released recently, demonstrating that 34.4\%$ \sim $48.5\% of the test cases expose the instability of LAPR techniques on average, showing the effectiveness of MT-LAPR and uncovering a positive correlation between code readability and the robustness of LAPR techniques. Inspired by the above findings, this paper uses the test cases generated by MT-LAPR as samples to train a CodeT5-based code editing model aiming at improving code readability and then embeds it into the LAPR workflow as a data preprocessing step. Extensive experiments demonstrate that this approach significantly enhances the robustness of LAPR by 49.32\% at most.

\end{abstract}


\section{Introduction}

As software systems become increasingly integral to daily life, the prevalence of software bugs rises accordingly. The widespread deployment of software systems means that bugs can result in system failures, security breaches, and diminished user experience. As such, developers often devote a significant amount of time and effort to bug fixing \cite{o2017debugging}. Automated Program Repair (APR) aims to alleviate this burden by automatically generating patches for the original buggy code, thereby minimizing the need for manual intervention by developers \cite{gazzola2018automatic}.

%As software programs and systems become more and more ubiquitous in everyday life, so do software bugs. Due to the wide-ranging adoption of software systems, software bugs can potentially lead to system failures, security vulnerabilities, and compromised user experience. As such, developers often need to spend a significant amount of time and effort to fix software bugs \cite{o2017debugging}. As a result, Automated Program Repair (APR) seeks to reduce the manual bug-fixing effort of developers by automatically synthesizing patches given the original buggy code \cite{gazzola2018automatic}. 

In recent years, various APR approaches have been investigated by researchers and practitioners to tackle the challenge of automatic bug fixing \cite{goues2019automated}.  The current state-of-the-art APR techniques predominantly revolve around Large Language Models (LLMs). Recent studies \cite{xia2023automated, xia2023plastic, kolak2022patch} have explored the application of LLMs in APR, demonstrating their powerful ability to repair software bugs, including those that were previously unrepairable by existing APR approaches. The impressive performance of LLMs has spurred the development of numerous LLM-powered Automated Program Repair (LAPR) tools and achieved promising results \cite{bouzenia2024repairagent, wei2023copiloting ,li2024hybrid}.

However, LLMs have been shown to be highly sensitive to input prompts \cite{shen2023chatgpt, wang2023robustness}. Semantically equivalent prompts with inconspicuous different expressions may lead LLMs to generate totally different outputs, causing instability in their performance. For example, Mastropaolo et al. \cite{mastropaolo2023robustness} studied the robustness of GitHub Copilot in code generation. The results show that modifying the prompt description results in different code recommendations in  $ \sim $ 46\% of cases. Also, differences in the semantically equivalent descriptions might impact the correctness of the generated code. Chen et al. \cite{chen2024nlperturbator}, through a series of experiments using six code LLMs for code generation, found that perturbed prompts can significantly decrease code generation performance, with reductions of up to 21.2\%. 
% Similarly, slight differences in the expressions of semantically equivalent programs potentially cause repair failures in some LLMs. 
Therefore, it is also crucial to conduct robustness testing on such LAPR techniques before their practical deployment, leaving the opportunity for remedy.  

% Previous studies proposed a series of approaches to evaluate the robustness of LLMs \cite{wang2021adversarial}, \cite{zhu2023promptbench}. 
% Although a series of studies have investigated the robustness of LLMs as above, they typically focus on perturbing prompts of natural language descriptions, such as sentiment analysis and code generation. 
% And these approaches are mostly based on random perturbations (e.g. the arbitrary replacement of characters in random positions.). 
Although several studies have investigated the robustness of LLMs, as mentioned above. Most of them typically focus on perturbing prompts of natural language descriptions, such as code generation \cite{mastropaolo2023robustness},\cite{chen2024nlperturbator} and sentiment analysis \cite{zhu2023promptbench}. However, programs with structural syntax and rigorous logic are more challenging to be perturbed while keeping the same semantics. Furthermore, perturbations for LAPR testing should reflect developers' diverse daily coding habits. Otherwise, the testing results will lack practical significance. Nevertheless, prevalent and practical perturbations on codes have not been studied and summarized yet. Motivated by the above,
% Considering the significance of developing the testing framework for the robustness of LAPR and no previous study has investigated in this area, 
this work takes an important step forward.
% Therefore, there is currently a significant lack of research on the robustness evaluation of LLMs in APR, highlighting the urgent need for a comprehensive evaluation framework.

In this paper, we propose MT-LAPR, a Metamorphic Testing framework for Large language model-powered Automated Program Repair techniques. 
% Specifically, to develop a comprehensive testing framework for automated program repair, we first need to understand what kind of difference real users might apply to actual code writing. Thus, we conduct a pilot study (Section \ref{3}) on 500 code samples collected from real users and a survey, and summarize nine metamorphic relations(MRs) across three perturbation levels:
% variable, statement, and block, making MT-LAPR provide MRs that reflect real-world user behaviors and are specially designed for automated program repair. 
Specifically, we first conduct a pilot study (Section \ref{3}) on 500 human-written code samples collected from Codeforces\cite{Codeforc98:online} to observe and collect real-world common coding discrepancies among developers, and then craft nine Metamorphic Relations (MRs) across three levels (i.e., token, statement, and block) for perturbation, making MT-LAPR provide MRs reflecting practitioners' real coding habits. MRs are designed algorithms implemented on Abstract Syntax Trees (ASTs) of buggy codes, thereby precisely locating and perturbing code elements. 
Subsequently, MT-LAPR employs these MRs on buggy codes from two extensively studied program repairing datasets, i.e., Defects4J and QuixBugs, to investigate the robustness of four recent LLMs (i.e., Mistral Large \cite{AuLargeM62:online}, LLaMA3-70/8B \cite{Llama311:online}, and CodeGemma-7B \cite{team2024codegemma}) in program repair. 
% that can be successfully repaired by LLM to generate test cases, which may mislead LLM and affect its program repair ability. 
% In our evaluation, we employ MT-LAPR to test four recent LLMs. 
Towards the Defects4J dataset, experimental results show that 40.1\%, 46.4\%, 56\%, and 54.7\% of the test cases expose the instability of Mistral Large, LLaMA3-70B, LLaMA3-8B, and CodeGemma-7B in program repair. As for the QuixBugs dataset, 4.9\%, 24.3\%, 33.2\%, and 75.2\% of the test cases reveal the repair failure after perturbation of LLMs in order, demonstrating the effectiveness of MT-LAPR. Besides, we also find a positive correlation between code readability and the performance of LAPR by continuously adding more perturbations to code samples. 

Inspired by the above finding, this paper uses the test cases generated by the MT-LAPR as samples to train a CodeT5-based code editing model aiming at improving code readability before feeding those buggy codes into LAPR techniques. Extensive experiments demonstrated that this approach significantly enhances the robustness of LAPR by 7.46\%$ \sim $49.32\%. %Codes, data, and results of our study in this paper are available at \cite{GitHuban89:online}. 
The main contributions of this paper are as follows:

%In this paper, we designed a comprehensive set of metamorphic testing methods to assess the program repair capabilities of large language models (LLMs) and developed a readability improvement model to improve the code repair performance of LLMs. Specifically, we first verified the prevalence and practical significance of perturbations in real-world scenarios through programmer experiments and surveys, summarizing nine common types of perturbations. Then, we constructed nine Java code perturbators based on Abstract Syntax Tree (AST), applied these perturbators to the selected dataset (defect4j), and designed a complete sampling method to obtain test cases. Subsequently, we conducted repair tests on these samples using large models to verify the impact of perturbations on the robustness of LLMs. Experimental evaluations demonstrated that all nine perturbators resulted in varying degrees of robustness degradation in LLMs. We further explored the changes in LLM robustness under different repair mode classifications. Finally, for the metamorphic testing of code, we developed a readability improvement model, JCRBooster, based on CodeT5. Evaluations showed that our model effectively improved the robustness of LLMs in program repair tasks. Specifically, the JCRBooster-Base model achieved improvements of 43.18\% and 7.46\% on LLaMA3-8B and LLaMA3-70B, respectively, while the JCRBooster-Large model achieved improvements of 49.32\% and 15.11\% on LLaMA3-8B and LLaMA3-70B, respectively. The overall working framework of this paper is shown in Figure~\ref{overview}.%

\begin{itemize}
    % \item The introduction of the first comprehensive testing framework, MT-LAPR, for LLM automated program repair robustness validation.
    \item We systematically construct the first metamorphic testing framework, namely MT-LAPR, to measure the robustness of LAPR techniques, proposing nine metamorphic relations on code across three perturbation levels that are widely accepted by practitioners.
    % \item A pilot study on 500 real-world code samples that leads to nine metamorphic relations, facilitating the implementation of MT-LAPR towards Java languages based on AST.
    % \item An extensive evaluation of MT-LAPR on four recent LLMs, demonstrating that MT-LAPR can generate perturbed codes that easily reduce the robustness of the four LLMs.
    \item Extensive experiments on MT-LAPR are conducted on four recent LLMs and two extensively examined bug-fixing datasets, demonstrating the effectiveness of MT-LAPR in detecting the instability of LAPR techniques. 
    % \item The concept of readability improvement is proposed based on common program metamorphic relation, and a readability improvement model is developed based on full parameter fine tuning pre-training model codeT5, which effectively improves the robustness of program repair for each LLM. 
    \item Based on the observation of the positive correlation between code readability and performance of LAPR techniques during perturbation, we propose a concept of improving code readability to confront the instability of LAPR and fine-tune a code readability improvement model with the above-generated test cases to testify our proposal.
\end{itemize}

% \begin{figure*}[h]
%   \centering
%   % \hspace*{-0.65cm}
%   \includegraphics[width=\linewidth]{overall.png}
%   \caption{MT-LAPR working framework}
%   \label{overview}
% \end{figure*}

\section{Background and Related Work}

\subsection{LLM-powered APR}

% For practitioners, fixing bugs is a time-consuming and labor-intensive task. In practice, developers must first figure out the problem and localize its root cause in the source code. Next, they speculate on various strategies to fix the issue and evaluate potential patches by applying them and checking if the test cases pass. If the tests fail, they conduct further debugging. Finally, they select the best patch and apply it to the codebase \cite{goues2019automated}. In this context, 
% Automatic Program Repair (APR) techniques have emerged to help developers automatically patch software bugs.
Automated Program Repair (APR) techniques are developed to help developers automatically patch software bugs. Recently, researchers have directly leveraged LLMs for various code-related tasks \cite{xue2024automated,yang2024exploring,xue2024escalating}, including APR, as they reduce the reliance on bug-fixing datasets and improve the accuracy and efficiency of bug fixes\cite{10172803}. LLM-powered APR (LAPR) is an innovative technique that utilizes pre-trained LLMs to automatically detect and fix software bugs. This approach leverages the extensive language understanding capabilities of LLMs to generate code patches that correct errors in software programs. Specifically, Xia et al.\cite{10172803} extensively explored the application of 9 state-of-the-art LLMs for APR. They evaluated LLMs in three repair settings and found that these models significantly outperformed existing APR techniques, with larger models generally achieving better results.  %On top of that, Xia et al. \cite{Xia2023ConversationalAP} proposed a novel approach called conversational APR, which utilizes LLMs for APR by integrating test case validation feedback in a conversational manner. 
Zhang et al. \cite{Zhang2022RepairingBI} suggested using an LLM trained on Codex to build an APR system named MMAPR, which can fix both syntactic and semantic errors in introductory Python programming assignments. %Li et al. \cite{li2024hybrid} introduced an innovative APR approach called GIANTREPAIR, which leverages the insight that LLM-generated patches and offers valuable guidance for the patch-generation process. 
Additionally, Kim et al. \cite{kim2022empirical} empirically investigated the performance of TFix in fixing errors from industrial Samsung Kotlin projects detected by a static analysis tool SonarQube. Matthew et al. \cite{jin2023inferfix} proposed InferFix, a transformer-based program repair framework, and deployed it as part of the Azure DevOps continuous integration pipeline internally at Microsoft. In summary, LLM-powered APR has been widely studied and applied in both academia \cite{bouzenia2024repairagent}, \cite{li2024hybrid}, \cite{Xia2023ConversationalAP, fu2022vulrepair, wu2023effective} and industry \cite{kim2022empirical}, \cite{jin2023inferfix}, \cite{joshi2023repair}, but corresponding testing systems are scarce, especially in robustness testing.
% In summary, LLM-powered APR represents a significant advancement in the field of software maintenance, offering a powerful and adaptable method for automating the repair of software bugs.

\subsection{Metamorphic testing}

%Ensuring the accuracy of program correctness is pivotal in software development. In software testing, a test oracle serves as a mechanism that compares the actual output of a program with the expected output to determine correctness and assess the success of the test. However, due to the complexity of verifying outputs in complex computations, the test oracle may become unavailable, leading to what is known as the test oracle problem \cite{zhou2004metamorphic}. %In such cases, testers often resort to manual prediction and verification of program outputs, which is a time-consuming and error-prone process. Therefore, the test oracle problem is recognized as one of the most challenging tasks in software testing \cite{zhou2004metamorphic}. To address this issue, Metamorphic Testing has been proposed. %Metamorphic testing is a technique used in software testing to generate new test cases from existing successful test cases, aiming to uncover software errors that may have been missed \cite{chen2020metamorphic}.
Metamorphic testing \cite{chen2020metamorphic} is a testing technique that aims to address the oracle problem \cite{zhou2004metamorphic}. It utilizes properties of the program, known as Metamorphic Relations (MRs), to generate follow-up test cases from existing successful test cases and automatically verify outputs without relying on an oracle \cite{segura2016survey}. Specifically, given an initial test case, metamorphic testing transforms it into a new test case using a pre-defined transformation rule and then checks whether the corresponding outputs of these test cases exhibit the expected relationships.
%Metamorphic testing involves three main steps \cite{segura2016survey}. First, it defines metamorphic relations to identify program properties and relationships among test inputs and outputs. Second, it generates or selects source test cases using methods like random testing. These cases are then used to apply the metamorphic relations. Finally, it executes the test cases, including follow-up cases generated from the relations, comparing their outputs to detect any failures that indicate the presence of bugs in the tested program. In conclusion, Metamorphic testing improves testing efficiency, reduces manual verification needs, and partially mitigates the test oracle problem.

In the past few years, metamorphic testing has been widely applied in various fields. Specifically, Wang et al. \cite{wang2023mttm} proposed a metamorphic testing framework for textual
content moderation software, which employs MRs on toxic textual content to generate test cases to evaluate textual content moderation software and algorithms. Chen et al. \cite{chen2009conformance} proposed the application of
metamorphic testing to test the conformance between network simulators and network protocols. In program-related areas, Chen et al. \cite{chen2002metamorphic} presented a case study of the application of metamorphic testing to programs implementing partial differential equations. Tao et al. \cite{tao2010automatic} selected the equivalence-preservation relation as the metamorphic relation and proposed an automatic metamorphic testing framework for the compiler. Xie et al. \cite{xie2013metamorphic}, \cite{xie2011spectrum} proposed the combination of metamorphic testing and Spectrum-Based Fault Localisation (SBFL) for debugging programs without an oracle. In this work, we, for the first time, adopt the methodology of metamorphic testing to assess the robustness of LAPR techniques.

\section{MT-LAPR}
\label{3}

This section first introduces a pilot analysis of code samples collected from real users. Then we introduce nine Metamorphic Relations (MRs) that are inspired by the pilot analysis, which can be
grouped into three categories: token-level, statement-level, and block-level. Finally, we discuss the MR combinations and test oracle automation. 

%In this section, we employ pilot analysis, including a programmer experiment and a survey, to explore whether code perturbation is common in real applications and whether it has practical significance. 
%This solves the problem: \textbf{RQ1: Whether the perturbation is common and has practical significance?}

\subsection{Pilot Analysis}
\label{a}

%\subsubsection{Data Source}

% In order to design comprehensive and precise perturbations and develop metamorphic relations(MRs),
In order to design comprehensive and practical perturbations, also called Metamorphic Relations (MRs),
we first conduct a pilot analysis on codes from real users to explore what kind of perturbations the users might apply to actual code writing. We consider code samples from Codeforces \cite{Codeforc98:online} for experiments, which is an online competitive programming platform recording a large number of problem-solving codes. 
% Therefore, we have collected a sufficient amount of real code samples submitted by programmers and systematically evaluated the readability of each individual submission. 

During the pilot study, we select 500 code samples from the ten problems featuring average difficulty and the highest passing rate, thereby ensuring the samples are representative and ample to find enough semantically consistent ones.
For each problem, fifty Java codes that passed the system tests are randomly selected as analysis samples. To facilitate understanding of the code and to identify differences between code written by different users, we select a relatively more readable code sample from those fifty submissions to serve as the benchmark sample. Then, two of the authors of this paper, each with 3-5 years of practical Java development experience, 
conduct a comparative analysis of the benchmark sample and the residual 49 sample codes for each coding problem, 
trying to identify the most prevalent different writing styles or structures. 
% focusing on code segments with consistent functionality. 
% They specifically aimed to identify the primary MRs present in the code samples from Codeforces. 
Finally, the two authors meet to discuss their findings, manually inspect all these code samples, and collectively summarize nine MRs that made the sample codes semantically equivalent but literally inconsistent with the benchmark code. 

% Accordingly, We identified nine MRs, each reflecting perturbation rules that real users might employ when writing code. 
We categorize these MRs from three levels: 1) token level, which primarily involves modifications to a single token, such as altering variable or method names within their scopes; 2) statement level, 
focusing on modifications of operators or operands within specific statements;
% focusing on expression alternations within specific statements, such as altering operators and operands. 
3) block level, involving adjustments to larger-scale code structures within specific blocks, such as equivalent transformations of loop structures or adding redundant code. 
% Such perturbations target the performance of code in different structures and contexts. 
Table \ref{AST} presents the nine perturbation rules, including their categories, examples, counts, and proportions, according to our pilot study. 
% sum and percentage. 
We introduce each MR in the following.

\begin{table*}[htbp]
\vspace{-1.5em}
\setlength{\abovecaptionskip}{0cm}
\renewcommand{\arraystretch}{1.3}
\caption{Summary of the perturbation categories in the pilot analysis}
\centering
\begin{tabular}{lllcc}
\toprule
\textbf{Perturbation Level}       & \textbf{Perturbation Rule} & \multicolumn{1}{c}{\textbf{Examples}}                                                              & \textbf{Counts} & \textbf{Proportion} \\ \hline
\multirow{2}{*}{Token Level}   & VariableRenaming             & int x = 1; → int y = 1;\ \  \ \     x++; → y++;                                                                       & 153          & 30.6\%              \\
                                  & MethodRenamimg               & int calculate() \{ ... \} → int compute() \{ ...  \}                                               & 104          & 20.8\%              \\ \hline
\multirow{3}{*}{Statement Level}   & AssignExpression             & x = x + y; → x += y;                                                                               & 107          & 21.4\%              \\
                                  & ConditionalExpression       & if (a \textgreater 0) → if (0 \textless a)                           & 91           & 18.2\%              \\
                                  & BinaryExpression             & int x = a + b; → int x = a - (-b);                                                                & 113          & 22.6\%              \\ \hline
\multirow{4}{*}{Block Level} & DummyVariable                & int x = 1; → int x = 1; int dummyVar = 2;                                                          & 54           & 10.8\%              \\
                                  & AddingComments               & int x = 1; → int x = 1; // Initialize x                                                            & 132          & 26.4\%              \\
                                  & VariableDeclaration          & void method() \{  ... ... int x;\}  →  void method() \{ int x;  ... ...  \}                                                           & 111          & 22.2\%              \\
                                  & ForToWhileLoop               & for (int i = 0; i \textless 10; i++)  \{  ... \}  → int i = 0; while (i \textless 10) \{  ...   i++; \} & 82           & 16.4\%              \\ \bottomrule
\end{tabular}
\label{AST}
\vspace{-1.5em}
\end{table*}

% \subsection{MRs with Variable-Level Perturbations}
\subsection{MRs with Token-Level Perturbations}
\label{b}

\noindent\textbf{MR1 VariableRenaming}

This MR replaces the name of an existing variable with another name, which involves the areas where it is defined and used. For example, a variable, namely $x$, may be replaced with $y$ as shown in Row 2 of Table \ref{AST}. %$\{int \ x = 1;\}\footnote{Here we use ``\{'' and ``\}'' to specify statements under certain scopes.}  \to \{int \ y = 1;\}$ and also in its usage area, such as $\{x++;\} \to \{y++;\}$. 
According to our statistics, MR1 is the most common literal inconsistency we find, accounting for 30.6\% of all samples, which also aligns with the fact that developers tend to use various naming conventions for variables \cite{gresta2023naming}. 
% This high frequency emphasizes the importance of using descriptive and meaningful variable names in the code.  
However, meaningless naming does not affect program semantics but damages the code understanding of developers and increases the likelihood of errors \cite{butler2010exploring,lavazza2023empirical}.  
% not only makes it difficult for other developers to understand the purpose and functionality of the code but also causes developers to spend more time on code inspection and debugging, delaying delivery and increasing the likelihood of errors \cite{lavazza2023empirical}.

\noindent\textbf{MR2 MethodRenaming}

This MR renames methods and updates all corresponding call sites within the repair context.
% within a specific scope. 
For example, developers may use the method name of $compute$ to replace $calculate$ for the definition of a computation-purpose function.
% In code writing, for the same method, different users may adopt different method names based on personal habits and judgment. For example, both \textit {calculate()} and \textit {compute()} can be used to define a method for computation. 
In some cases, users might not adhere to any specific naming convention and instead use arbitrary names, which undoubtedly increases the difficulty of understanding and maintaining the code \cite{hofmeister2017shorter}. 
% This MR is useful for evaluating the code's reliance on specific method names, ensuring that functionality is maintained regardless of naming conventions. It also plays a role in code obfuscation, making it more challenging for developers to reverse engineer the code. 
%It is worth noting that method names generally have contextual calling dependencies. If only the method names within a block are changed, it can disrupt the context's dependency relationships. In such cases, even if an LLM successfully repairs a buggy block, the method invocation relationships within the class no longer exist, leading to program errors. Therefore, after applying the MR9 perturbation, we performed a reverse process on the code repaired by the LLM, replacing the modified method names back to their original names. This was done to ensure a more objective and fair evaluation of the robustness of the LLM.

\subsection{MRs with Statement-Level Perturbations}
\label{c}

\noindent\textbf{MR3 AssignExpression}

This MR modifies assignment expressions, which changes direct assignments (i.e., ``='') to compound assignment operators (e.g., ``+='', ``-='', etc.), such as Row 4 of Table \ref{AST}. Using compound assignment operators is sometimes not as clear as direct assignment but is semantically equivalent and increases code conciseness.
% Such modification does not alter the program's behavior but changes the appearance and semantic representation of the code. 
In our pilot study, 21.4\% of the samples contain different forms of assignment expressions with the same semantics. 
% While this operation can increase code conciseness, it also increases the difficulty for programmers to quickly understand the code to a certain extent.

\noindent\textbf{MR4 ConditionalExpression}

This MR alters simple conditional expressions, often by changing the comparison operator, such as Row 5 of Table \ref{AST}.
%$\{if(a>0);\} \to \{if(0<a);\}$.
Conditional expressions are core elements of control flow in programming languages \cite{green1977conditional}, and different forms of conditional expressions are widely used in practical coding, accounting for 18.2\% as counted in our pilot study. Different representations of the same conditional expression can affect the readability of the code \cite{holst2021importance}, thereby influencing the extent to which developers understand the code.

\noindent\textbf{MR5 BinaryExpression}

This MR changes the operator/operand in binary expressions but keeps the equivalent semantics, as an example shown in Row 6 of Table \ref{AST}, %$\{int \ x = a + b;\} \to \{int \ x = a - (-b);\}$, 
adding a number is equivalent to subtracting its negative. 
% This modification aids in testing code behavior with different operators or checking for dependencies on specific operators. 
This MR, which accounts for 22.6\% of all samples, may not only prolong code review and debugging times but also increase the likelihood of misinterpretation and errors by programmers in practice.

\subsection{MRs with Block-Level Perturbations}
\label{d}

\noindent\textbf{MR6 DummyVariable}

This MR inserts declarations of dummy variables into the code. These variables serve no practical purpose in code logic and are redundant in the codebase. For example, as shown in Table \ref{AST}, variable $dummyVar$ is declared but not utilized, introducing unnecessary redundancy into the code. 
During practical code writing, some programmers may inadvertently leave behind variable declarations that are no longer used when making adjustments to the code. While these redundant codes do not affect the functionality of the code, they can diminish its readability, thereby increasing the difficulty for users to comprehend.  

%To someone reading the code, the addition of dummy variables makes it take more time to distinguish which variables are actually involved in the computation and which are just "noise".

\begin{comment}
\begin{figure}[h]
  \centering
  % \hspace*{-0.65cm}
  \includegraphics[width=\linewidth]{eg1.png}
  \caption{An example of DummyVariable}
  \label{eg1}
\end{figure}
\end{comment}

\noindent\textbf{MR7 AddingComments}

This MR adds single-line comments to the code that are not relevant to the context of the code or have no effect on the logic. Notably, in our pilot study, 132 samples had comments that interfered with reading, accounting for 26.4\% of the total. Although the purpose of comments is to enhance understanding of the code, poor quality or irrelevant comments can have the opposite effect, leading to programmers misinterpreting or misunderstanding the code's functionality.

%At the core of this strategy is the addition of elaborate comments to seemingly insignificant parts of the code that appear to be closely related to the surrounding code, but which do not contribute to the logic, and are only intended to make it more difficult for the onlooker to understand.

\noindent\textbf{MR8 VariableDeclaration}

This MR moves variable declarations to the beginning of their corresponding scopes. In the writing of the same code, different programmers may declare a desired variable in different places. Although such a difference does not affect program semantics as these variables are placed in the same scope before usage, some of them may hinder program comprehension to some extent. According to our pilot study, 111 samples, accounting for 22.2\% of the dataset, occur in such a situation.
% had issues with variable declarations. 
% This MR highlights the importance of consistent scoping and storage, as variable declarations in different locations can lead to data leakage and increased maintenance challenges.

%This is typically done to adhere to coding conventions or to test if the code relies on specific variable scopes.

\noindent\textbf{MR9 ForToWhileLoop}

This MR converts for-loops into equivalent while-loops. The for-loop and while-loop are the two most common types of loop structures, and they can be equivalently transformed \cite{wolf1991loop}. Consequently, in practice, different programmers, or even the same programmer at different locations within the same code, may use different loop structures for coding. For example, as shown in Table \ref{AST}, the for-loop structure is changed to a while-loop structure.

\subsection{Implementation Details}
\label{detail}
This section presents the implementation details of each MR we proposed. Given a Java function, we first exert JavaParser\cite{JavaPars88:online} 
% (an open-source project library for parsing, analyzing, and modifying Java code \footnote{https://javaparser.org/}) 
to parse the Java code string to obtain the AST, thereby traversing and localizing AST nodes to make corresponding perturbations. Considering a coding style may be replicated many times within one's code in practice \cite{ogura2018bring}, we transform all encountered AST nodes that satisfy our perturbation criteria in each sample to mimic such a situation. 
%This section presents the implementation details of each MR we proposed. Given a Java function, we first exert JavaParser to parse the Java code string to obtain the Compilation Unit, thereby traversing and localizing AST nodes to make corresponding perturbations.

\noindent\textbf{MR1 VariableRenaming} 

For each encountered variable declaration node, a new variable name is generated, which consists of the original variable name and a dynamic suffix. This method aims to preserve as much of the original code information as possible, avoiding the performance damage of LLMs owing to information loss instead of perturbations.  
The perturbator maintains a variable name mapping table to record the correspondence between the original variable name and the new variable name. When encountering variable usage nodes during the traversing, the perturbator updates the variable name according to the mapping table. The perturbation logic is as follows:
\begin{equation}
\setlength{\abovedisplayskip}{4pt}
varName \rightarrow concat(varName, ``\_var", C)
\setlength{\belowdisplayskip}{4pt}
\end{equation}
where $varName$ is the original variable name, $``\_var"$ is a fixed infix, $C$ is an incrementing counter to discriminate different variables within the same function, $concat(\cdot)$ denotes the concatenation of all its parameters sequentially, and $\to$ denotes the transformation symbol connecting before- and post-perturbation.

\noindent\textbf{MR2 MethodRenamimg} 

%For each method declaration node, except the main method and constructors, it generates a new method name by appending a counter to the original name and updates the method declaration. The perturbator also updates method names in all corresponding call sites with the new method names. The perturbation logic is as follows:
For each method declaration node, the perturbator similarly updates the original method name by appending a dynamic suffix. It is noticeable that main methods and constructors are excluded because the former is directly invoked by the JVM to initiate program execution, and altering its name would prevent the program from starting correctly, while the latter's name must match their corresponding class names. The perturbator also updates method names in all corresponding call sites accordingly within the repair context.  The perturbation logic is as follows:
% \[ \text{$methodName$} \rightarrow \text{$methodNameMethod$$C$} \]
\begin{equation}
\setlength{\abovedisplayskip}{3pt}
 methName \rightarrow concat(methName, ``Method", C)
 \setlength{\belowdisplayskip}{3pt}
\end{equation}
where $methName$ is the original method name, $``Method"$ is a fixed infix, and $C$ is an incrementing counter.

\noindent\textbf{MR3 AssignExpression} 
% For each method declaration node, the modifier identifies 

This MR towards each identified assignment expression node that assigns the result of a binary expression to a target variable. If the target of the assignment is one of the operands of the binary expression, the perturbator replaces the original assignment with a corresponding compound assignment (e.g., Row 4 of Table \ref{AST}). The perturbation logics are as follows:
\begin{equation}
\setlength{\abovedisplayskip}{3pt}
\begin{aligned}
target=target \ op \ expr; \rightarrow target \ op \ = \ expr; \\
target=expr \ op \ target; \rightarrow target \ op \ = \ expr;
\end{aligned}
\setlength{\belowdisplayskip}{3pt}
\end{equation}
where $target$ is the target variable of the assignment, $op$ is a binary operator (e.g., +, -, *, /), and $expr$ is the non-target operand of the binary expression.

\noindent\textbf{MR4 ConditionalExpression} 

For each encountered binary expression node, if the operator is ``$>$'', it swaps the operands and changes the operator to ``$<$''. If the operator is ``$<$'', it swaps the operands and changes the operator to ``$>$''. For boolean expressions with ``$\&\&$'' or ``$||$'', it adds extra parentheses around the expression. The perturbation logics are as follows:
\begin{equation}
\setlength{\abovedisplayskip}{3pt}
 a > b \rightarrow b < a,~a < b \rightarrow b > a,~ a \ \text{$op$} \ b \rightarrow (a \ \text{$op$} \ b)
 \setlength{\belowdisplayskip}{3pt}
\end{equation}
where \text{$op$} is either ``\&\&'' or ``$||$''.
% , and ``,'' is used for separation among different rules.

\noindent\textbf{MR5 BinaryExpression} 

For each encountered binary expression node, it applies different transformations based on the operator. If the operator is ``+'' or ``*'', it swaps the operands. If the operator is ``-'', it changes the operation to the addition of the negated right operand. If the operator is ``/'', it changes the operation to multiplication by the reciprocal of the right operand. String concatenation operations are excluded from this perturbation. The perturbation logics are as follows:
\begin{equation}
\setlength{\abovedisplayskip}{3pt}
\begin{aligned}
a + b \rightarrow b + a, &~~ a * b \rightarrow b * a \\
a - b \rightarrow a + (-b), &~~ a / b \rightarrow a * (1 / b)
\end{aligned}
\setlength{\belowdisplayskip}{3pt}
\end{equation}
where \(a\) and \(b\) are the operands of the binary expression.

\noindent\textbf{MR6 DummyVariable} 
% For each method declaration node, a new dummy variable declaration is added to the method's body. 

Dummy variable declarations are inserted as sub-trees under method declaration nodes with random positions among their sibling nodes.
The dummy variable here is an integer ``0'' with a name composed of a fixed prefix, i.e., ``dummyVar'', followed by an incrementing counter for discrimination. 
% The modifier maintains a counter to generate unique variable names.
% Each dummy variable is inserted at a random position within the method's body. 
The inserted statement of this perturbation can be formally defined as follows:
% \[MethodBody \rightarrow MethodBody~\text{with}~ int~dummyVarC = 0; \]
\begin{equation}
\setlength{\abovedisplayskip}{3pt}
 int \ concat(``dummyVar", C) \ =0;
 \setlength{\belowdisplayskip}{3pt}
\end{equation}
% where 
% $dummyVarC$ is the new dummy variable name, $dummyVar$ is a fixed prefix, and 
% $C$ is an incrementing counter, while others are fixed strings to be composed for the statement.

\noindent\textbf{MR7 AddingComments} 
% For each method declaration node, a unique line comment is added to the method's body.

Similar to MR6, this MR designates single-line comments to be inserted as sub-trees under method declaration nodes with random positions among their sibling nodes. 
The comment content is a fixed string (i.e., ``//This method was modified -'') followed by a randomly generated $UUID$ \cite{green1997unique}. 
% The new comment is inserted at a random position within the method's body. 
The inserted statement of this perturbation can be formally defined as follows:
% \[ MethodBody \rightarrow \\
% MethodBody~\text{with}~comment + UUID \]
\begin{equation}
 % concat(comment, UUID) 
 \setlength{\abovedisplayskip}{3pt}
 concat(``//This \ method \ was \ modified \ \text{-}", UUID)
 \setlength{\belowdisplayskip}{3pt}
\end{equation}
% where $comment$ is ``This method was modified -'' and $UUID$ is randomly generated by method \emph{java.util.UUID.randomUUID()}.

\noindent\textbf{MR8 VariableDeclaralion} 

% For each method and constructor declaration node, this perturbator moves variable declarations to the top of the body. It traverses the AST nodes within the method or constructor body, identifying variable declaration expressions. 
Within the scope of methods, for each encountered variable declaration node, if it includes an initializer, its declaration and initialization should be separated, where the declaration is moved to the top while the assignment is placed in its original position. The perturbation logic is as follows:
\begin{equation}
\setlength{\abovedisplayskip}{3pt}
type~var~=~value; \rightarrow type~var; \ ... \ var~=~value;
\setlength{\belowdisplayskip}{3pt}
\end{equation}
where $type$ is the type of the variable, $var$ is the variable name and $value$ is the value of the variable.
%where variable declarations without initializations are moved to the top, and declarations with initializations are split into separate declaration and assignment statements.

\noindent\textbf{MR9 ForToWhileLoop} 
% For each encountered `for` statement node, it constructs a new `while` loop with the same loop condition. 

For each encountered for-loop sub-tree, the perturbator transforms it to a corresponding sub-tree of the while-loop.
The initialization expressions (e.g., ``int i=0;'' in Row 10 of Table \ref{AST}) are extracted and placed before the while-loop, and the update expressions (e.g., ``i++;'' in Row 10 of Table \ref{AST}) are moved to the end of the while-loop body. The perturbation logic is as follows:
\begin{equation}
\setlength{\abovedisplayskip}{3pt}
\begin{split}\text{$for(init; cond; update) \{ body \}$} \rightarrow \\\text{$init; while(cond) \{ body; update; $\}}\end{split}
\setlength{\belowdisplayskip}{3pt}
\end{equation}
where $init$ are the initialization expressions, $cond$ is the loop condition, $update$ are the update expressions, and $body$ is the original loop body.

\subsection{Discussion}
\label{discussion}
\noindent\textbf{Combinations of different MRs.} 
As evidenced by the Column \textbf{Proportion} in Table \ref{AST}, the sum of the proportion is over 100\%, as one code sample normally carries more than one different writing style against the benchmark sample, which means it is necessary to inject multiple perturbations in mutant constructions, thereby simulating the practical situation.
% it should be noted that one code sample may carry more than one positions with different writing styles against the bechmark sample, which means it is mecessary to inject multiple perturbarions to construct test cases, thereby simulating the practical situation.
MT-LAPR allows for the combination of different MRs. By layering varying numbers of MRs, we can generate diverse mutants.
% test cases\footnote{The test case construction process is described in \ref{EM}}. 
Afterward, we define perturbation distance ($pd$), representing the number of different MRs applied in generating a mutant, which can be calculated by the following formula:
\begin{equation}
\setlength{\abovedisplayskip}{3pt}
pd = d(P,P')=\sum_{n=1}^{9}\delta (m_{n} ,P,P')
\label{cal}
\setlength{\belowdisplayskip}{3pt}
\end{equation}
where $P$ is the original code. $P'$ is the code after applying one or more MRs. $m_n$ is the $n$-th MR.
$\delta(m_{n} ,P,P')$ is an indicator function defined as:
\begin{equation}
% \delta (m_{n} ,P,P') =\begin{cases}
%   1,& \text{if } m_n \ is\ applied \ to\ generate\ P' \\
%   0,& \text{if } m_n \ is \ not\ applied
\setlength{\abovedisplayskip}{3pt}
\delta (m_{n} ,P,P') =\begin{cases}
  1,& \text{if } P \neq P' \ s.t. \ m_n \\
  0,& \text{if } P = P'
\end{cases}
\setlength{\belowdisplayskip}{3pt}
\label{cal1}
\end{equation}

For example, an original code \( P \) is subjected to a combination of perturbations using MR1, MR4, MR5, and MR9, resulting in the perturbed code \( P' \). According to eq. (\ref{cal})-(\ref{cal1}), the perturbation distance (\( pd \)) can be calculated as \( pd = 4 \).
% As such, we can explore the impact of perturbations of multiple combined MRs on LAPR.

\noindent\textbf{Automating test oracle procedure.} Given the test inputs have been automatically mutated based on the above MRs, here we introduce the test oracle automation process of MT-LAPR to complete the constitution of test cases in this work. Considering the mutants are semantically equivalent, we adopt testing suites\footnote{test suites are provided by bug-fixing datasets for the correctness evaluation of repaired programs, which should be discriminated from test cases generated by MT-LAPR in this paper.} of original samples to re-evaluate the correctness of the repair results of their corresponding mutants, where a repair is considered successful if the whole test suite passes without any error.  
It is worth noting that method/variable names typically have contextual dependencies, and testing suites require original method/variable names for execution. Simply changing variable/method names within specific code snippets may disrupt these contextual dependencies and cause the testing suite to fail to locate them. In such cases, even if LLMs can successfully repair a buggy code snippet, the disrupted calling relationships would still lead to program errors. Therefore, before evaluating the repair results, we first perform reverse processing on the perturbed code repaired by LLMs towards variable/method names altered by \textbf{MR1} and \textbf{MR2}, thereby ensuring a more accurate evaluation of the LLM's robustness.

\section{Study Design}
\label{4}

This section details the design of specific experiments under the MT-LAPR framework, including studied models, dataset construction, evaluation metrics, and evaluation methodology.

\subsection{Studied Models}
\label{Studied Models}
In this paper, we choose four recent LLMs: Mistral Large, LLaMA3-70B, LLaMA3-8B, and CodeGemma-7B. 

\begin{itemize}
  \item Mistral Large \cite{AuLargeM62:online}: A newly released closed-source model from Mistral AI. This model learns the intrinsic patterns of language through pre-training tasks and is fine-tuned using labeled data to enhance performance on specific tasks. Currently, Mistral Large has achieved excellent results in common benchmark tests \cite{mcdonald2024reducing}. 
  \item LLaMA3 \cite{Llama311:online}: A newly released family of multilingual large language models employing a decoder-only transformer architecture offers configurations with 8 billion and 70 billion parameters. LLaMA3 is pre-trained on over 15 trillion tokens from publicly available sources \cite{huang2024good}. We include the two versions of 8B and 70B for experiments.
  \item CodeGemma \cite{team2024codegemma}: 
  A collection of specialized open code models built on top of Gemma \cite{team2024gemma},
  which is based on transformer decoder architecture crafted by Google. CodeGemma is further trained on more than 500 to 1000 billion tokens of primary code and offers configurations with 2 billion and 7 billion parameters. We select the version of 7B for experiments.
\end{itemize}

The parameter settings and prompt templates for all LLMs reviewed are fixed the same.
Particularly we set \textit{temperature}=0, by which LLMs will only select tokens with the highest probability when inference \cite{peeperkorn2024temperature}. As such, we can control LLMs to produce identical outputs under the same context and conditions, eliminating any randomness and attributing all variations of outputs to code perturbations. 
% This ensures that changes in the LLM's repair results for perturbed samples are solely due to the perturbation itself, and not influenced by the LLM's inherent randomness.

\subsection{Dataset Construction}
\label{Dataset Construction}
We select the two most widely examined program repair datasets for experiments, including Defects4J and QuixBugs.
Defects4J \cite{just2014defects4j} is a dataset containing 395 real bugs (version 1.2) collected from six open-source Java projects. Specifically, each bug contains a buggy version and a fixed version, as well as a corresponding test suite that triggers that bug for patch validation. QuixBugs \cite{lin2017quixbugs} contains 40 single-line Java bugs from 40 programs. It is a dataset in the lab, different from the Defects4J collecting bugs from the real world.

Since we use different combinations of perturbation rules to craft mutants from the original buggy code, the number of corresponding test cases can be extremely large. To conclude a valuable conclusion in a reasonably limited time, we sample a subset of original buggy codes as base samples for experiments. Besides, we only consider samples that can be successfully repaired by LLMs as base samples. Because, for repair-failed samples before perturbation, once it still cannot be repaired after perturbation, it is hard to determine whether the perturbation affects LLMs' repair or just due to LLMs' inherent limitations.
In this case, we first require each LLM with the parameter settings mentioned in Section \ref{Studied Models} to repair every sample in each dataset and filter out those successfully repaired samples. 
Subsequently, to ensure a comprehensive and representative sampling, we follow Sobreira et al. \cite{sobreira2018dissection}'s taxonomy on buggy code to select samples from each LLM's repair result, where the taxonomy divides buggy code into nine categories according to repair patterns. In selecting samples, we aim to cover all repair patterns as much as possible. Ultimately, we obtain 60 base samples from each dataset, where 15 samples for each LLM, namely Defects4J$_{base}$ and QuixBugs$_{base}$. 

\subsection{Evaluation Metrics}
\label{Evaluation Metrics}
To evaluate how well MT-LAPR does in generating test cases that affect LLMs, i.e., the robustness of LLMs under test, we propose the evaluation metric, namely $R$-$score$. It calculates the proportion of test cases that the LLM successfully repairs out of the total number of test cases, 
which can be formally defined as:
\begin{equation}
\setlength{\abovedisplayskip}{3pt}
R\text{-}score=\frac{valid}{valid + invalid}
\setlength{\belowdisplayskip}{3pt}
\end{equation}
where $valid$ represents the number of test cases in which the LLM's repair is successful, and $invalid$ represents the opposite. The lower the $R\text{-}score$, the better the testing framework in detecting the instability of LLMs. Note that before testing with MT-LAPR, LLMs have the $R\text{-}score$=1 on both Defects4J$_{base}$ and QuixBugs$_{base}$.

%\subsubsection{AST Construction}
%We implement the nine perturbations described in section \ref{3} based on AST. AST(Abstract Syntax Code) is an intermediate representation of programming language source code during compilation or interpretation, constructed through lexical and syntactic analysis. It presents the code structure in a tree-like structure, where each node represents a meaningful unit in the code, such as variables, operations, control structures, etc., and the connections between the nodes reflect their logical relationships. AST ignores non-critical details in the source code, such as spaces and comments, which allows the compiler to efficiently carry out the subsequent code optimization and analysis work.

\subsection{Evaluation Methodology}
\label{EM}

To evaluate the effectiveness of MT-LAPR, we propose five Research Questions (RQs) below:

\begin{comment}
\begin{itemize}
  \item RQ2: Can perturbations reduce the robustness of LLMs?
  \item RQ3: Are robustness and  number of perturbation combination negatively correlated?
  \item RQ4: Are there better perturbations and useless perturbations?
  \item RQ5: Does the repair performance of LLMs differ across various repair patterns?
\end{itemize}
\end{comment}

\subsubsection{RQ1: Whether the perturbations are prevalent and practically significant?}
\label{rq1}
In Section \ref{3}, we propose nine different MRs through a pilot analysis and design various perturbation rules for each MR. This RQ aims to explore whether these perturbations are prevalent and practically significant in real-world scenarios. To this end, we design a survey. Specifically, we invite ten full-time Java developers (at least 3-5 years of coding experience for each) from the industry to serve as survey respondents.
% and give them a questionnaire, which contains nine questions, each corresponding to a type of perturbation. 
Given a questionnaire, they are required to choose on a five-point Likert scale \cite{bertram2007likert} and answer how frequently they encounter each of our proposed perturbations in practice based on their experience (1 = Never, 5 = Very Frequent). 
%How severe are the losses or issues likely caused by these types of perturbations? (1 = Negligible, 5 = Severe). 
To ensure respondents can understand each perturbation, practical examples are provided next to each question.

%\subsubsection{RQ2: Can perturbations affect the robustness of LLMs?}

% \subsubsection{RQ2: Can MT-LAPR identify deficiencies of LLMs in robustness for APR?}
\subsubsection{RQ2: Can MT-LAPR identify robustness deficiencies of LLMs in APR?}
This RQ aims to investigate the overall robustness of each recent LLM examined and evaluate the effectiveness of our testing framework. By exploring the ability of LLMs to repair buggy code after perturbation, practitioners can learn how much they can trust an LLM under test. 

\subsubsection{RQ3: How perturbation distance affects the performance of LLM-powered APR?}
Pilot analysis (Section \ref{3}) already demonstrates that the occurrence of multiple MRs within the same code segment is a common phenomenon in practice. Therefore, this RQ explores the impact of the number of different MRs (i.e., perturbation distance) on the performance of LAPR.

\subsubsection{RQ4: Are there better perturbations and useless perturbations?}
This RQ aims to discuss the influence of every single-distance perturbation we propose on identifying the robustness deficiency of LAPR, thereby exploring which one is more important in robustness testing on LAPR techniques. 

%We manually designed and implemented nine different code perturbations, however, there may be cases where some of the perturbations are of low utility, and these so-called “useless perturbations” may not have a significant impact on evaluating the robustness of LLM. Therefore, we analyze how effective each perturbation is.

\subsubsection{RQ5: Buggy codes of which repair patterns are (not) prone to be affected by perturbations?}
Although the previous RQ has thoroughly investigated various MRs, at the samples' angle, different types of buggy code in LAPR may react differently to perturbations. For instance, bug-fixing for samples under certain repair patterns might be more resistant to perturbations, while others may be more sensitive to them. Therefore, this RQ focuses on studying these phenomena.

In order to answer RQ2$\sim$5 systematically and comprehensively, we employ perturbation rules of both single-distance ($pd$=1) and multi-distance ($pd$=2$\sim$9) to construct test cases. For each dataset (i.e., Defects4J$_{base}$ and QuixBugs$_{base}$), let the base sample set for the \(i\)-th LLM be \(S_i = \{s_{ij}\}\) with \(j\) representing the \(j\)-th base sample, where \(i = \{1, 2, 3, 4\}\) and \(j = \{1, 2, \ldots, 15\}\) in our experiments. The perturbation rules are defined as \(M = \{m_n\}\) with \(n\) representing the \(n\)-th perturbation rule, where \(n = \{1, 2, \ldots, 9\}\) in our experiments. To examine whether each perturbation is successfully applied, we use the Python module \textit{difflib} \cite{ragkhitwetsagul2016similarity} to compare the differences between $P$ and $P'$. Simultaneously, for each base sample \(s_{ij}\), we maintain a corresponding perturbation list: \(L_{ij} = \text{$List$}[m_k]\), where \(m_k \subseteq M\), denoting the perturbation rules \(m_k\) that successfully perturb the \(j\)-th sample of the \(i\)-th LLM.

% In constructing single-distance perturbation test cases, we applied the 9 perturbation rules to the 60 base samples individually. To ensure each perturbation was successfully applied, we used a $diff$ tool to compare the differences between the samples before and after perturbation, verifying the successful injection of the perturbations. Based on this process, we obtained the single-distance perturbation test cases derived from each base sample. Simultaneously, for each base sample \(s_{ij}\), we generated a corresponding perturbation list: \(L_{ij} = \text{List}[m_k]\), where \(m_k \subseteq M\). This denotes the perturbation rules \(m_k\) that successfully perturbed the \(j\)-th sample of the \(i\)-th LLM.

In constructing single-distance perturbation test cases, we apply the 9 perturbation rules to the 60 base samples of each dataset individually.
For multi-distance perturbation test cases, we use a combinatorial sampling method. For a given perturbation distance ($pd$) and the perturbation list \(L_{ij}\) of the base sample \(s_{ij}\), the number of test cases \(Num_{ij}^{pd}\) can be generated from \(s_{ij}\) is calculated below:
\begin{equation}
\setlength{\abovedisplayskip}{3pt}
{Num_{ij}^{pd} = {\binom{l}{pd}} = {\frac{l!}{pd!(l-pd)!}}} , 2\le pd\le l
\setlength{\belowdisplayskip}{3pt}
\end{equation}
where \(l\) represents the length of \(L_{ij}\). This allows us to derive multi-distance perturbation test cases from each base sample. For instance, if the first base sample of the first LLM (\(s_{11}\)) can be perturbed by \(m_1\), \(m_3\), and \(m_5\), i.e., \(L_{11} = [m_1, m_3, m_5]\), then the total number of multi-distance perturbation test cases it can generate is 4=$Num_{11}^{2}+Num_{11}^{3}=3+1$. To balance the number of test cases across LLMs and control the computational overhead, if \(Num_{ij}^{pd} > 20\), we randomly select 20 test cases. Based on the above procedures, we construct test cases from Defects4J$_{base}$ and QuixBugs$_{base}$, eventually obtaining Defects4J$_{test}$ and QuixBugs$_{test}$.% and the detailed statistic information is listed in \cite{GitHuban89:online}.
%to avoid exponential growth in the number of test cases.

Finally, we evaluate the repair performance of the four LLMs on all test cases. For RQ2, we calculate the repair performance of each LLM on all test cases to determine their robustness via $R\text{-}score$ proposed in Section \ref{Evaluation Metrics}. For RQ3, gathering the repair results of all LLMs under test together, we assess their average performance variation in program repair at different perturbation distances. For RQ4, focusing on single-distance perturbation, we calculate the average performance of all LLMs on test cases of each single-distance perturbation and discuss their influence from the perspectives of perturbed contents and scales. For RQ5, partitioning test cases by repair patterns according to their corresponding base samples, we study the average performance of all LLMs in program repair, thereby exploring the sensitivity to perturbations of buggy code in diverse categories when bug-fixing.

\section{Results and discussion}

\subsection{RQ1: Whether the perturbations are prevalent and practically significant?}

\begin{figure}[h]
\vspace{-1em}
\setlength{\abovecaptionskip}{0cm}
  \centering
  % \hspace*{-0.65cm}
  \includegraphics[width=\linewidth]{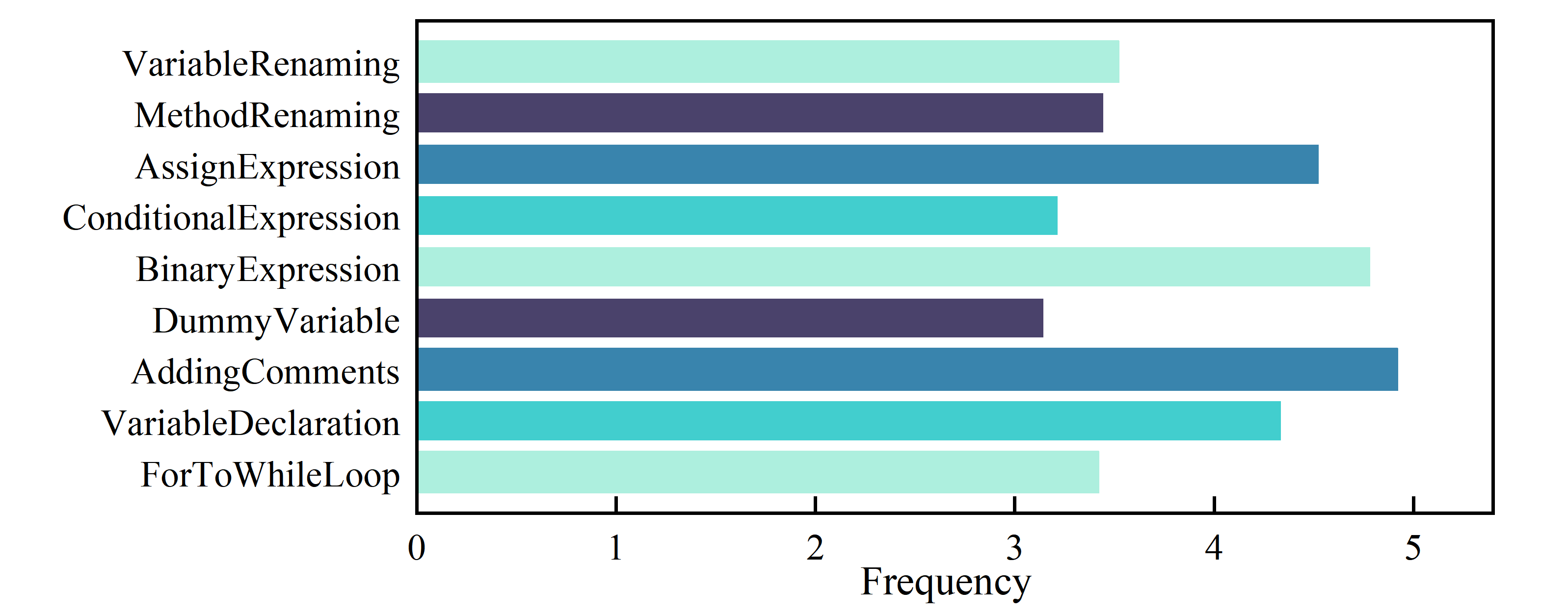}
  \caption{The results of the survey}
  \label{5}
\vspace{-1.5em}
\end{figure}

Figure \ref{5} presents the overall results of the questionnaire survey, which offered programmers' subjective judgments on the frequency of the nine specific code perturbations. We measure the inter-rater agreement using Randolph’s Kappa coefficient \cite{kirk2021hatemoji}, obtaining a value of 0.76, which indicates ``almost perfect agreement''. According to the survey results, the average frequency scores of the nine perturbations are all greater than 3 (Generally Frequent), indicating that these perturbations are prevalent in practice and hold practical significance.

%Based on the analysis of the figure, we can categorize the perturbations into three groups: high frequency and high severity, high frequency but low severity, and high severity but low frequency. First, the perturbations with high frequency and high severity include BinaryExpression and MethodRenaming. BinaryExpression is quite common and can significantly impact program correctness, while MethodRenaming, although less common, can cause method call errors and severely affect program functionality. Second, the perturbations with high frequency but low severity are AddingComments and ForToWhileLoop perturbator. AddingComments is a common practice, but incorrect comments typically do not affect code execution. ForToWhileLoop occurs frequently, but has a relatively low severity score, indicating that the error has a minor impact.
%Lastly, the perturbations with high severity but low frequency include VariableRenaming, VariableDeclaration, AssignExpression, and ConditionalExpressions. These perturbations may not occur as frequently, but they can significantly impact code readability, and program logic, and even lead to data and permission leakage when they do arise.

%In conclusion, the analysis demonstrates that different perturbations have varying levels of impact on code quality and program functionality. It is essential to be aware of the potential issues each perturbation can cause, with a particular focus on those that are both frequent and severe, as they pose the greatest risk to the integrity and maintainability of the codebase.

\begin{boxK}
  \faIcon{pencil-alt}  \textbf{Answer to RQ1:} Through a survey, we prove that the nine perturbations are prevalent and practically significant.
\end{boxK}

%\subsection{RQ2: Can perturbations affect the robustness of LLMs?}
% \subsection{RQ2: Can MT-LAPR identify deficiencies in LLM robustness for APR?}
\subsection{RQ2: Can MT-LAPR identify robustness deficiencies of LLMs in APR?}
\label{5B}
%Table \ref{tab:rotllm} presents the repair performance of four LLMs (LLaMA3-8B, LLaMA3-70B, CodeGemma-7B, Mistral Large) on the test cases. We observe that the robustness of the LLaMA3-8B, LLaMA3-70B, CodeGemma-7B, and Mistra-7B models are 0.536, 0.440, 0.453, and 0.599, respectively, with an overall robustness of 0.515. The robustness of each LLM has declined to varying degrees, all falling below 0.6. This indicates that the test cases generated by our MT-LAPR framework can effectively inject perturbations into the code and successfully impact the performance of LLMs in program repair.

Table \ref{tab:rotllm} presents the repair performance of four LLMs (i.e., Mistral Large, LLaMA3-70B,  LLaMA3-8B, and CodeGemma-7B) on test cases from two datasets. We observe that the $R\text{-}score$ of Mistral Large, LLaMA3-70B, LLaMA3-8B, and CodeGemma-7B on Defects4J$_{test}$ is 0.599, 0.536, 0.440, and 0.453, respectively, with an overall $R\text{-}score$ of 0.515. On QuixBugs$_{test}$, the $R\text{-}score$ are 0.951, 0.757, 0.668, and 0.248, respectively, resulting in an overall $R\text{-}score$ of 0.656. The robustness of each LLM has decreased to varying extents, demonstrating that the test cases generated by our MT-LAPR framework can successfully impact the LLMs' code repair performance. Additionally, the relatively better performance of LLMs on QuixBugs$_{test}$ might be attributed to the typically simpler structure of the bugs in QuixBugs, which are often classic algorithm problems that are easier to understand and repair.

Among the four LLMs tested, LLaMA3-8B and CodeGemma-7B show the largest performance declines across both datasets. This indicates that these two models, due to their smaller parameter sizes, have poorer perturbation resistance in APR. In contrast, the robustness performance of LLaMA3-70B is better than that of LLaMA3-8B, suggesting that the scaling effect exists for APR, where larger models tend to achieve better performance \cite{xia2023automated}. Additionally, we observe that Mistral Large performs the best among the four LLMs, surpassing LLaMA3-70B. Although the specific parameter size of Mistral Large is not disclosed, the model demonstrates excellent perturbation resistance in APR.

\begin{comment}
\begin{table}[htbp]
\caption{Robustness of each LLM under perturbation}
\centering
\begin{tabular}{cccc}
\toprule
\textbf{Model} &\textbf{Invalid} &\textbf{Valid} &\textbf{Robustness}   \\ \midrule
\multirow{1}*{LLaMA3-70B} & 355 & 410 & 0.536  \\
\multirow{1}*{LLaMA3-8B} & 497 & 390 & 0.440  \\
\multirow{1}*{CodeGemma-7B} & 296 & 245 & 0.453  \\
\multirow{1}*{Mistral Large} & 405 & 604 & 0.599  \\
\midrule
\multirow{1}*{Total Robustness} & \multicolumn{3}{c}{0.515} \\
\bottomrule
\end{tabular}
\label{tab:rotllm}
\end{table}
\end{comment}

\begin{table}[htbp]
\vspace{-1.5em}
\setlength{\abovecaptionskip}{0cm}
\caption{Robustness evaluation of each LLM under test}
\centering
\renewcommand{\arraystretch}{1.25}
\resizebox{\linewidth}{!}{
% \begin{threeparttable}
\begin{tabular}{lcccccc}
\toprule
\multicolumn{1}{c}{\multirow{2}{*}{\textbf{Model}}} & \multicolumn{3}{c}{Defects4J\textsubscript{test}} & \multicolumn{3}{c}{{QuixBugs\textsubscript{test}}}                                                                            \\ \cmidrule(lr){2-4} \cmidrule(lr){5-7}
\multicolumn{1}{c}{}                                & \multicolumn{1}{l}{\textbf{Invalid}} & \multicolumn{1}{l}{\textbf{Valid}} & \multicolumn{1}{l}{\textbf{R-score}} & \multicolumn{1}{l}{\textbf{Invalid}} & \multicolumn{1}{l}{\textbf{Valid}} & \multicolumn{1}{l}{\textbf{R-score}} \\ \hline
Mistral Large                                       & 405                                  & 604                                & 0.599                                   & 103                                  & 1982                               & 0.951                                   \\
LLaMA3-70B                                          & 355                                  & 410                                & 0.536                                   & 480                                  & 1497                               & 0.757                                   \\
LLaMA3-8B                                           & 497                                  & 390                                & 0.440                                   & 693                                  & 1392                               & 0.668                                   \\
CodeGemma-7B                                            & 296                                  & 245                                & 0.453                                   & 1568                                 & 517                                & 0.248                                   \\
 \hline
\textbf{Avg. R\text{-}score}                                    & \multicolumn{3}{c}{0.515}                                                                                           & \multicolumn{3}{c}{0.656}                                                                                           \\ \bottomrule
\end{tabular}}
\label{tab:rotllm}
\vspace{-1.5em}
\end{table}

\begin{boxK}
  \faIcon{pencil-alt}  \textbf{Answer to RQ2:} The test cases generated by our MT-LAPR framework can effectively identify the robustness deficiencies of LLM in automated program repair.
  % impact the performance of LLMs in program repair.
\end{boxK}

\subsection{RQ3: How perturbation distance affects the performance of LLM-powered APR?}
\label{rq3result}

\begin{figure}[h]
\vspace{-1em}
\setlength{\abovecaptionskip}{0cm}
  \centering
  % \hspace*{-0.65cm}
  \includegraphics[width=\linewidth]{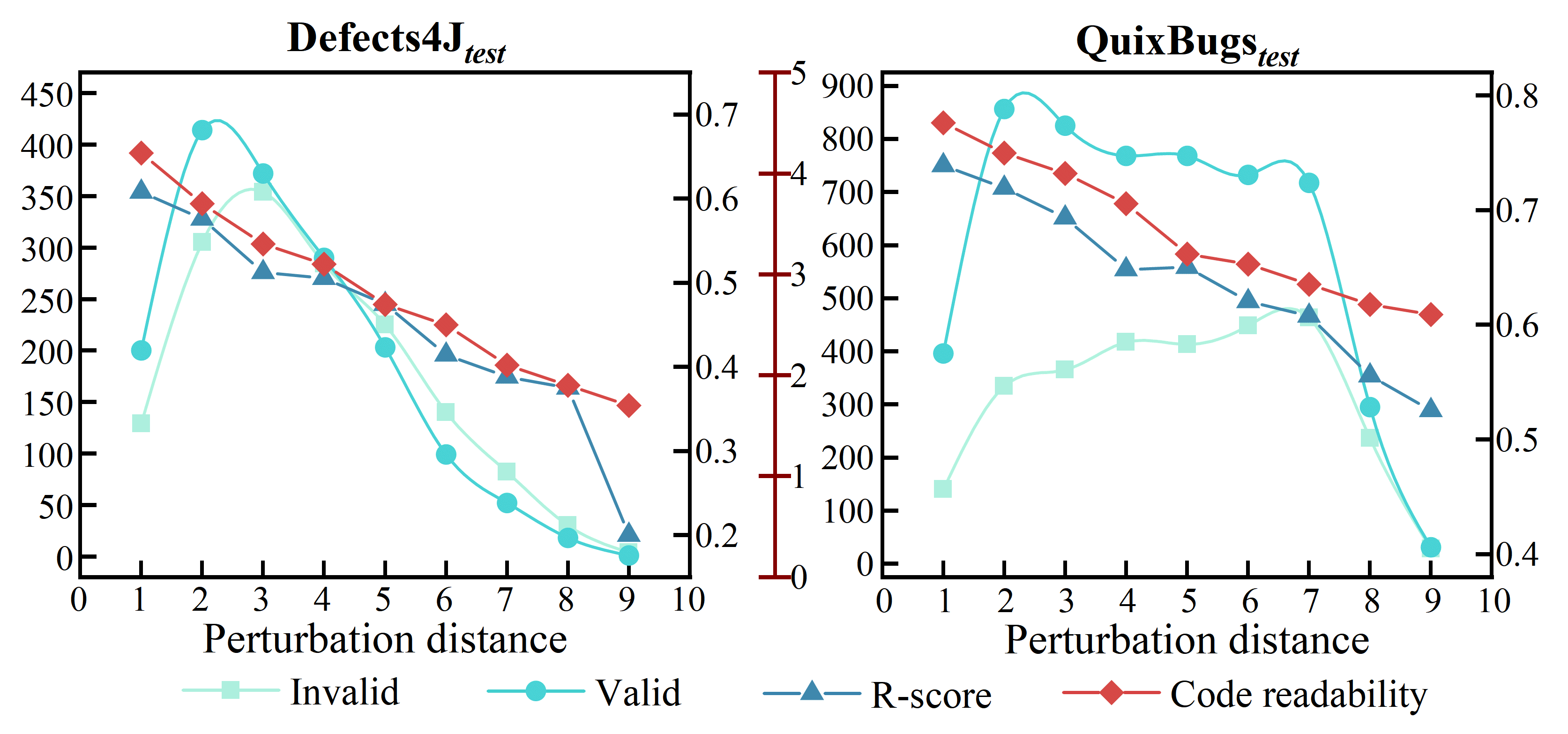}
  \caption{R-score and code readability vary with the perturbation distance}
  \setlength{\abovecaptionskip}{0.1cm}
  \label{result3}
\vspace{-1.5em}
\end{figure}

Figure \ref{result3} shows the average performance trends ($R\text{-}score$ lines) across all LLMs as the perturbation distance increases, where the trends are the same from both the Defects4J$_{test}$ and QuixBugs$_{test}$:   
% It can be observed that the LLMs exhibit the same trend on test cases perturbed from both the Defects4J$_{base}$ and QuixBugs$_{base}$: 
as the perturbation distance increases, the $R\text{-}score$ of the LLMs gradually decreases, indicating a negative correlation between perturbation distance and LLMs' performance. An explanation is as the perturbation distance increases, the perturbations become more complex and varied, leading to a greater extent of code obfuscation \cite{sarker2024syntactic}. Besides, some of the MRs proposed in Section \ref{3} result in damage to code comprehension, perturbing code with more MRs also probably decreases code readability, potentially deteriorating LLMs' performance in program repair. As such, more robustness deficiencies are found by our generated test cases.

%Specifically, we select a base sample from both Defects4J$_{base}$ and QuixBugs$_{base}$, ensuring that both samples could be perturbed by all nine perturbation rules. We then select nine semantically equivalent test cases from Defects4J$_{test}$ and QuixBugs$_{test}$ for each base sample, with each test case representing a different perturbation distance. Subsequently, we distribute a questionnaire to 10 programmers as described in \ref{rq1}. The questionnaire is divided into two parts based on the dataset source of the test cases, with each part providing the nine test cases mentioned above. 
To verify the above hypothesis, we further conduct a survey. Specifically, we select five base samples from Defects4J$_{base}$ and QuixBugs$_{base}$, respectively, ensuring that all chosen samples could be perturbed by all nine perturbation rules. For each dataset, we then select nine semantically equivalent test cases corresponding to the chosen five base samples, each representing a different perturbation distance ($pd=1\sim9$). 
% Based on this, we design five questionnaires, where each 
% for the different base samples. Each questionnaire 
% is divided into two parts according to datasets and provides nine test cases selected from the corresponding dataset towards the corresponding base samples. 
Subsequently, for each dataset, ten developers, as described in RQ1 (Section \ref{rq1}), are invited and divided into pairs to evaluate the code readability of each base sample and its corresponding mutated test cases on a five-point Likert scale (1 = Very Unreadable, 5 = Very Readable) \cite{dorn2012general} based on their judgment.
% and each group was given a different questionnaire.
% Respondents are asked to independently rate the code readability for each test case on a five-point Likert scale (from ``Very Unreadable” 1 to ``Very Readable” 5) \cite{dorn2012general} based on their judgment. 
Additionally, the concept of code readability is provided for programmers: \emph{Code readability is the amount of mental effort required to understand the code} \cite{sedano2016code}. To validate the effectiveness of the assessment, we compute Cohen’s kappa coefficient \cite{chmura2002kappa} to assess the consistency between the two programmers of each group. In our survey, the overall mean kappa scores on each dataset
% A and B 
are 0.65 and 0.67, respectively, indicating that they have reached a ``substantial agreement" and that our assessment process is effective. Finally, we compute the average readability score for test cases at different perturbation distances, and the results (red lines) are illustrated in Figure \ref{result3}. As the perturbation distance increases, the readability of the corresponding test cases gradually decreases. This indicates a negative correlation between code readability and perturbation distance.
% Simultaneously, due to the lack of descriptiveness and the potential presence of redundant code after perturbation, the overall readability of the code decreases, making reverse engineering more difficult. Therefore, the increase in perturbation distance enhances the interference with the LLMs, resulting in decreased robustness.

%Figure. \ref{result3} illustrates the trend of LLM robustness as the number of perturbation combinations increases. It is evident that the larger the number of perturbation combinations, the robustness decreases gradually, indicating a negative correlation between the number of perturbation combinations and robustness. This suggests that as the the number of perturbation combinations increases, the perturbations become more complex and diverse, leading to increased code obfuscation. Concurrently, due to the lack of descriptiveness and the potential presence of redundant code after perturbation, the overall readability of the code declines, making reverse engineering more difficult. Therefore, the increase in the number of perturbation combinations intensifies the interference with the LLM, resulting in a decrease in robustness.

\begin{boxK}
  \faIcon{pencil-alt}  \textbf{Answer to RQ3:} There is a significant negative correlation between the performance of LAPR and perturbation distance, as posing more MR-based perturbations on buggy codes damages their readability.
\end{boxK}

\subsection{RQ4: Are there better perturbations and useless perturbations?}

% Considering that the edited contents and the number of edited tokens both vary among different perturbations, we present Table~\ref{tab:result4} to illustrate the average editing distances and $R$-$score$s of all LLMs examined grouped according to different perturbation rules. 
Table~\ref{tab:result4} shows the testing results grouped according to different perturbation rules.
% revealing the impact of each single-distance perturbation on the identifying robustness deficiencies of LAPR. 
It can be observed that all nine perturbation rules individually uncover the instability of LAPR more or less, proving each of their effectiveness in the MT-LAPR framework. On Defects4J$_{test}$, the perturbation with the greatest impact on the LLMs' performance is \textit{ConditionalExpression}, potentially because it determines the core logic of control flows, affecting the main functionality of a code snippet more.  %This perturbation modifies conditional expressions, which can significantly change the appearance of the code, thus affecting the repair ability of LLMs. On the other hand, 
Besides, the perturbation with the least impact on the LLMs' performance is \textit{AddingComments}, as program compiling will not consider comments and by no means affect program logic. Thus, potentially, LLMs hardly take action toward comment-related altering.  %This perturbation only adds single-line comments to the code without changing its meaning, thus having minimal impact on LLMs. 
In contrast, on QuixBugs$_{test}$, the $R$-$score$ of LLMs to various perturbation rules is relatively balanced, with a range of only 0.085. Potentially because QuixBugs only consists of lab code 
% is smaller in scale (the lines of code per program ranging from 9 to 67 lines \cite{ye2021comprehensive}) and 
with single-line bugs, lacking the complex context dependencies. Therefore, the impact of different perturbation rules on the test cases in QuixBugs$_{test}$ is relatively limited and more evenly distributed. 

%Additionally, to explore the impact of the number of operations in each perturbation on the robustness performance of the LLM, we calculated the $ edit\  distance$ \cite{levenshtein1966binary} between the perturbed code and the original code. 

\begin{table}[htbp]
\vspace{-1.5em}
\setlength{\abovecaptionskip}{0cm}
\caption{Effectiveness of single-perturbations}
\centering
\renewcommand{\arraystretch}{1.25}
\resizebox{\linewidth}{!}{
\begin{tabular}{lcccccccl}
\toprule
\multirow{2}{*}{\textbf{Perturbation Rule}} & \multicolumn{1}{l}{\multirow{2}{*}{\textbf{Edit Distance}}} & \multicolumn{3}{c}
{Defects4J\textsubscript{test}} & \multicolumn{3}{c}{{QuixBugs\textsubscript{test}}}
             & \multicolumn{1}{c}{\multirow{2}{*}{\textbf{Avg.}}} \\ \cmidrule(lr){3-5} \cmidrule(lr){6-8}
                                            & \multicolumn{1}{l}{}                                        & \multicolumn{1}{l}{\textbf{Invalid}} & \multicolumn{1}{l}{\textbf{Valid}} & \multicolumn{1}{l}{\textbf{R-score}} & \multicolumn{1}{l}{\textbf{Invalid}} & \multicolumn{1}{l}{\textbf{Valid}} & \multicolumn{1}{l}{\textbf{R-score}} & \multicolumn{1}{c}{}                               \\ \hline
VariableRenaming                            & 17.09                                                       & 16                                   & 25                                 & 0.610                                   & 16                                   & 44                                 & 0.733                                   & 0.683                                              \\
MethodRenamimg                              & 6.39                                                        & 25                                   & 35                                 & 0.583                                   & 15                                   & 45                                 & 0.750                                   & 0.667                                              \\ \hline
AssignExpression                            & 9.76                                                        & 2                                    & 3                                  & 0.600                                   & 14                                   & 45                                 & 0.763                                   & 0.750                                              \\
ConditionalExpressions                      & 11.67                                                       & 17                                   & 17                                 & \textbf{0.500}                                   & 13                                   & 46                                 & 0.780                                   & 0.677                                              \\
BinaryExpression                            & 11.73                                                       & 10                                   & 11                                 & 0.524                                   & 18                                   & 41                                 & 0.695                                   & 0.650                                              \\ \hline
DummyVariable                               & 10.09                                                       & 27                                   & 33                                 & 0.550                                   & 15                                   & 45                                 & 0.750                                   & 0.650                                              \\
AddingComments                              & 11.42                                                       & 16                                   & 44                                 & 0.733                                   & 18                                   & 42                                 & 0.700                                   & 0.717                                              \\
VariableDeclaration                         & 11.42                                                       & 10                                   & 22                                 & 0.688                                   & 16                                   & 44                                 & 0.733                                   & 0.717                                              \\
ForToWhileLoop                              & 13.39                                                       & 6                                    & 10                                 & 0.625                                   & 15                                   & 44                                 & 0.746                                   & 0.720                                              \\ \bottomrule
\end{tabular}
}
\label{tab:result4}
\vspace{-1em}
\end{table}

Apart from the edited content, perturbations also vary in edited tokens, which also may affect LLMs' performance.
% Additionally, we explore the impact of the number of tokens changed in each perturbation on the robustness performance of LLMs. 
To this end, we calculate the average edit distance \cite{levenshtein1966binary}  between the perturbed code and the original code across all LLMs examined, with the results shown in Table~\ref{tab:result4}. 
Apparently, \textit{VariableRenaming} has the largest edit distance, owing to the presence of multiple variable declarations and calls within a single buggy code snippet that satisfies the perturbing criteria.
% , resulting in a higher number of tokens being altered. 
On the other hand, the perturbation with the least edit distance is \textit{MethodRenaming}. Although this perturbation is also at the token level, a buggy code snippet generally contains only one method and a limited number of corresponding call sites, hence the number of tokens changed is minimal. However, both perturbations significantly impact the LLM's $R$-$score$. Therefore, it is hard to summarize any patterns between edit distances and LLMs' performance, which is the same situation with other perturbations. Thus, we speculate that the number of edited tokens has negligible impacts on LLMs' performance. 
% We use a robustness score of 0.7 as the threshold, where a score greater than 0.7 indicates that the perturbation had a minor impact on the repair performance of the LLMs, and a score less than 0.7 indicates a significant impact. We then calculate the average edit distance for test cases with robustness scores greater than and less than 0.7, which are 11.498 and 11.394, respectively, showing almost no difference. This suggests that the number of token changes may have a weak correlation with the robustness performance of LLMs. 
To further verify this conjecture, we calculate the Spearman correlation coefficient \cite{myers2014spearman} between edit distances and $R$-$score$s. The results show a correlation coefficient of 0.042 and a $p$-value of 0.914$>$0.05, indicating no significant relationship between the edit distance and the performance of LLMs. 

\begin{boxK}
  \faIcon{pencil-alt}  \textbf{Answer to RQ4:} The nine perturbation rules designed in MT-LAPR are all effective in instability detection. Perturbation content is more significant than the number of perturbed tokens. 
\end{boxK}
% under review
\subsection{RQ5: Buggy codes of which repair patterns are (not)
prone to be affected by perturbations?}

% Based on the repair patterns classifications described in Section \ref{4}, 
As mentioned in Section \ref{4}, base samples are selected across various repair patterns to ensure the comprehensiveness of our experiments. Here we present Table \ref{tab:result5} to illustrate how buggy codes of different repair patterns react to perturbations when repairing with LLMs.  
% present the robustness performance of LLMs under different repair patterns. 
%In all test cases, the proportion of samples corresponding to each repair pattern remains consistent with the original proportions \cite{sobreira2018dissection}. This indicates the balance of our test case distribution and confirms the validity of our sampling method. However, 
Repair patterns with ``-'' on the $R$-$score$ field denote that we do not include any corresponding samples of certain datasets, as no LLM tested in this work can successfully repair them, thereby cannot be involved in the base sample set as mentioned in Section \ref{Dataset Construction}. 
% Our test cases do not include any samples corresponding to the \textit{Copy/Paste} repair pattern, as no LLMs tested in this work can successfully repair such samples.
% This pattern involves copying code from one location and pasting it into another to fix missing or duplicated code fragments. This type of repair typically requires complex contextual understanding and precise localization of code fragments. Due to the token limit of LLMs and their limited ability in structural and semantic understanding of code \cite{wei2023copiloting}, LLMs find it challenging to successfully repair samples of this pattern.

%两个数据集上修复模式的表现有些对立。下面这一段是原来针对Defects4J写的
In the performance of LLMs under other repair patterns, the $R$-$score$ in the \textit{Constant Change} category is 0.583, the highest in Defects4J$_{test}$. This repair pattern involves changing fixed values in the program, usually to fix hardcoded values or incorrect constants. This suggests that LLMs perform well when handling simple value modifications, as constant changes typically have minimal impact on code logic and are relatively easy for the model to identify and repair. It is noteworthy that in this pattern, the LLM's robustness performance on QuixBugs is the worst, with only 0.086. This is because all test cases for this repair pattern in QuixBugs originate from one project named ``SQRT", and this sample could only be repaired by CodeGemma-7B. Therefore, the experimental result here is highly contingent, and we ignore it.

%Additionally, the robustness in the Expression Fix category is 0.543. This pattern includes correcting errors in mathematical operations, logical operations, or other complex expressions. This indicates that the model has a good ability to understand and correct expressions, with less impact from perturbations.

In contrast, the $R$-$score$ of the two datasets under the \textit{Missing Null-Check} category is 0.298 and 0.357, which is the lowest among all repair patterns. This repair pattern is related to the addition of conditional expressions or the expansion of existing ones with null checks. This shows that after introducing perturbations, LLMs perform poorly in handling missing null checks, possibly because accurately identifying potential null pointer risks and adding appropriate checks involves complex contextual understanding. Similarly, LLMs perform poorly under the \textit{Wraps-with/Unwraps-from} repair pattern, which involves wrapping existing code with a conditional branch. This repair pattern requires LLMs to accurately understand the usage scenarios and logic of the wrapping structure, involving complex structural adjustments. Therefore, perturbing these code samples leads to more difficulties for LLMs in bug-fixing.

\begin{comment}
\begin{table}[htbp]
\caption{Robustness in various repair patterns}
\centering
\begin{tabular}{cccc}
\toprule
\textbf{Repair Patterns} &\textbf{Invalid} &\textbf{Valid} &\textbf{Robustness}   \\ \midrule
\multirow{1}*{Code Moving} & 55 & 76 & 0.420  \\
\multirow{1}*{Conditional Block} & 990 & 1236 & 0.445  \\
\multirow{1}*{Constant Change} & 140 & 100 & 0.583  \\
\multirow{1}*{Copy/Paste} & 0 & 0 & -  \\
\multirow{1}*{Expression Fix } & 1013 & 851 & 0.543  \\
\multirow{1}*{Missing Null-Check} & 158 & 373 & 0.298  \\
\multirow{1}*{Single Line} & 718 & 650 & 0.525  \\
\multirow{1}*{Wraps-with/Unwraps-from} & 418 & 640 & 0.395  \\
\multirow{1}*{Wrong Reference} & 572 & 591 & 0.492  \\
\multirow{1}*{Total} & 2562 & 2434 & 0.513  \\
\bottomrule
\end{tabular}
\label{tab:result5}
\end{table}
\end{comment}

\begin{table}[htbp]
\vspace{-1.5em}
\setlength{\abovecaptionskip}{0cm}
\caption{Robustness of bug-fixing for codes in various repair patterns}
\centering
\renewcommand{\arraystretch}{1}
\resizebox{\linewidth}{!}{
\begin{tabular}{lcccccc}
\toprule
\multirow{2}{*}{\textbf{Repair Patterns}} & \multicolumn{3}{c}{\textbf{Defects4J$_{test}$}}                                                                              & \multicolumn{3}{c}{\textbf{QuixBugs$_{test}$}}                                                                               \\ \cmidrule(lr){2-4} \cmidrule(lr){5-7} 
                                          & \multicolumn{1}{l}{\textbf{Invalid}} & \multicolumn{1}{l}{\textbf{Valid}} & \multicolumn{1}{l}{\textbf{R-score}} & \multicolumn{1}{l}{\textbf{Invalid}} & \multicolumn{1}{l}{\textbf{Valid}} & \multicolumn{1}{l}{\textbf{R-score}} \\ \hline
Code Moving                               & 55                                   & 76                                 & 0.420                                   & 109                                  & 169                                & 0.608                                   \\
Conditional Block                         & 990                                  & 1236                               & 0.445                                   & 635                                  & 1311                               & 0.674                                   \\
Constant Change                           & 140                                  & 100                                & 0.583                                   & 127                                  & 12                                 & 0.086                                   \\
Copy/Paste                                & 0                                    & 0                                  & -                                       & 0                                    & 0                                  & -                                       \\
Expression Fix                            & 1013                                 & 851                                & 0.543                                   & 2346                                 & 3909                               & 0.625                                   \\
Missing Null-Check                        & 158                                  & 373                                & 0.298                                   & 268                                  & 149                                & 0.357                                   \\
Single Line                               & 718                                  & 650                                & 0.525                                   & 632                                  & 1592                               & 0.716                                   \\
Wraps-with/Unwraps-from                   & 418                                  & 640                                & 0.395                                   & 0                                    & 0                                  & -                                       \\
Wrong Reference                           & 572                                  & 591                                & 0.492                                   & 352                                  & 1486                               & 0.808                                   \\ \bottomrule
\end{tabular}
}
\label{tab:result5}
\vspace{-1.5em}
\end{table}

\begin{boxK}
  \faIcon{pencil-alt}  \textbf{Answer to RQ5:} Buggy codes of different categories significantly affect the repair capabilities of LLMs, where easier ones are more perturbation-resistant while more difficult ones are more sensitive.
  % LLMs perform well when dealing with simple constant modifications and single-line repairs but perform poorly when handling complex structural adjustments and repairs that require deep contextual understanding. 
\end{boxK}

\section{Robustness Improvement}
RQ3 (Section \ref{rq3result}) has proved the positive correlation between buggy code readability and LLMs' APR performance. Therefore, improving code readability has a great probability of making LLMs' APR performance robust. As such, we propose our solution of training a code editing model to improve the code readability of existing programs.    
% \subsection{Improvement method}
\subsection{Methodology and setup}
Code editing models are trained to construct a mapping ($f$) from less readable codes ($X$) to readable codes ($X'$), namely $X' = f(X)$. Considering we have already designed a series of MRs that can perturb codes to less readable statuses, we adopt pre-perturbed code as $X'$ while applying MRs to $X'$ to construct $X$. To mitigate the threat of data leakage, we still test on the dataset used in previous experiments (RQ2$\sim$5), while preparing the training dataset with the rest of the samples. Due to the limited number of remaining samples in QuixBugs, it is difficult to generate sufficient training samples. Hence, we focused on constructing the model for Defects4J. Ultimately, we obtain a training dataset consisting of 30,471 $\left <x,x' \right>$ pairs, where $x \in X$ and $x' \in X'$.

Considering the powerful generality of Pre-trained Language Models (PLMs) in recent years, we use a widely examined PLM, i.e., CodeT5 for experiments \cite{fu2022vulrepair}, \cite{liu2024delve}, \cite{liu2023empirical}. Both CodeT5-base (220M) \cite{wang2021codet5} and CodeT5-large (770M) \cite{le2022coderl} are experimented for fine-tuning on the prepared training samples, where inputs are the perturbated codes ($X$) and ground truths are corresponding pre-perturbation codes ($X'$). In experiments, we train each model with 3 epochs, a learning rate of 5 × \(10^{-5}\), a batch size of 1, and a weight decay of 0.01. As such, we obtain two code editing models, namely CodeT5-base$^\star$ and CodeT5-large$^\star$, for improving code readability.

We embed our fine-tuned models into the LAPR workflow as a data pre-processing step to improve the buggy code readability before feeding them to LLMs for repair. We evaluate their effectiveness on two LLMs (i.e., LLaMA3-8B and LLaMA3-70B) tested before, considering the computational overhead and time efficiency.

\subsection{Results}
Table \ref{table5} showcases the two LLMs' program repair performance before and after improving buggy code readability via CodeT5-base$^\star$ and CodeT5-large$^\star$. Apparently, equipped with CodeT5-base$^\star$, LLaMA-8B obtains an improvement by 43.18\% in terms of $R$-$score$, while LLaMA-70B achieves an improvement by 7.46\%. Moreover, when incorporating CodeT5-large$^\star$ into the workflow of LAPR, LLaMA-8B and LLaMA-70B achieve more significant improvements by 49.32\% and 15.11\% in terms of $R$-$score$, respectively.
To delve deep into the editing effectiveness of both models in improving code readability, we further investigate the edit distances from perturbed buggy code and edited ones to original ones. We find that both CodeT5-base$^\star$ and CodeT5-large$^\star$ reduce the edit distances of around 75\% of perturbed codes, where the former can edit 17.18\% of samples to their original status while the latter can do the same for over 20\% samples, indicating our models can effectively improve code readability and further lift the robustness of LAPR while bypassing the access of their inner parameters.  
% The experimental results are shown in Table \ref{table5}. The results of JCRBooster-Base and JCRBooster-Large are both improved, for LLaMA-70B, the improvement is 7.46\% and 15.11\%, for LLaMA-8B, the improvement is 43.18\% and 49.32\%, respectively.  
%In \ref{rq3result}, we have demonstrated that code readability decreases as perturbation distance increases. 
%To validate whether JCRBooster successfully improves code readability, we select samples that could be successfully repaired by LLM after being processed by JCRBooster. We then calculate the average edit distance between the samples before and after JCRBooster processing, compared to the original undisturbed samples. The results show that the average edit distance for the samples before JCRBooster processing was 17.554. In contrast, after JCRBooster processing, the average edit distance decrease to 4.998. Therefore, our approach can significantly enhance the robustness of LAPR and effectively improve code readability.

\begin{table}[htbp]
\vspace{-1.5em}
\setlength{\abovecaptionskip}{0cm}
\caption{Robustness Improvement Results}
\centering
\renewcommand{\arraystretch}{1.7}
\resizebox{\linewidth}{!}{
{\fontsize{25}{20}\selectfont
\begin{tabular}{lccccccccc}
\toprule
\multirow{2}{*}{\textbf{Model}} & \multicolumn{3}{c}{\textbf{Perturbed}}                   & \multicolumn{3}{c}{\textbf{Edited by CodeT5-base$^\star$}}                                                            & \multicolumn{3}{c}{\textbf{Edited by CodeT5-large$^\star$}}                                                           \\ \cmidrule(lr){2-4} \cmidrule(lr){5-7} \cmidrule(lr){8-10} 
                                & \textbf{Invaild} & \textbf{Vaild} & \textbf{R-score} & \textbf{Invaild} & \textbf{Vaild} & \textbf{R-score}                                                 & \textbf{Invaild} & \textbf{Vaild} & \textbf{R-score}                                                 \\ \hline
LLaMA3-8B                       & 497              & 390            & 0.440               & 328              & 559            & \textbf{\begin{tabular}[c]{@{}c@{}}0.630\end{tabular}} & 304              & 583            & \textbf{\begin{tabular}[c]{@{}c@{}}0.657\end{tabular}} \\
LLaMA3-70B                      & 355              & 410            & 0.536               & 324              & 441            & \textbf{\begin{tabular}[c]{@{}c@{}}0.576\end{tabular}}  & 293              & 472            & \textbf{\begin{tabular}[c]{@{}c@{}}0.617\end{tabular}} \\ 
% \cmidrule(lr){1-4} \cmidrule(lr){5-7} \cmidrule(lr){8-10}
% Edit Distance                   & \multicolumn{3}{c}{-}                                   & \multicolumn{3}{c}{17.554 → \textbf{4.998(71.53\%↓)}}                                                             & \multicolumn{3}{c}{18.010 → \textbf{5.124(71.55\%↓)}}                                                            
\bottomrule
\end{tabular}
}
}
\label{table5}
\vspace{-1.5em}
\end{table}

% \begin{boxK}
%   \faIcon{pencil-alt}  \textbf{Answer to RQ6:} By constructing code editing model JCRBooster, test cases generated by MT-LAPR can enhance the robustness of LAPR and improve code readability.
% \end{boxK}

\subsection{Trail and error}

There are various methods that could potentially improve the robustness of LLMs. For example, one approach is to retrain the examined LLMs. However, this method might impair the existing capabilities of the LLM. Moreover, given the enormous number of parameters in LLMs, this approach poses a significant challenge for practical implementation. Alternatively, we attempt to have LLMs perform code refactoring on buggy codes for readability improvement via a direct inference manner. This method aims to retain the essential logic in buggy code while removing unnecessary information before being fed for APR. However, our experiments show that this approach led to a 20.5\% reduction in the $R$-$score$ of LLMs, indicating that LLMs struggle to refactor code without adequate training. Additionally, designing reverse rules \cite{sarker2024syntactic} might improve code readability. However, in practice, the specific perturbation rules of codes are diverse and infinite, while designing each rule manually would consume substantial effort and restrict scalability. Our fine-tuning method, bypassing manual efforts, can effectively overcome the above limitations, allowing LLMs to enhance their generality through supervised learning without tuning their inner parameters.

\section{Threats to Validity}

% We consider the following threats to the validity.

\noindent\textbf{Internal Validity.}
%There are a few possible internal threats to validity within our work. 
First, the nine MRs introduced herein may fail to encapsulate the entire spectrum of coding style discrepancies encountered in real-world development, thereby restricting the generality of experimental outcomes. To mitigate this threat, we adopt a holistic workflow encompassing user behavior analysis, metamorphic relations synthesis, test case creation, and supervised fine-tuning to evaluate and improve the robustness. 
% In the advent of novel circumvention tactics, 
As such, people can follow the above workflow to devise fresh MRs. Second, focusing exclusively on the Java Programming Language (PL) might limit the generality of the results. However, Java is one of the most widely used PLs in practical development. Furthermore, we conduct our tests on two extensively recognized datasets in the APR field, making the experimental outcomes more representative.
% , i.e., Defects4J and QuixBugs, which significantly mitigates this threat. 
Future studies can extend our findings by incorporating additional datasets from other PLs and domains.
Finally, since the datasets we used have been widely studied, data leakage may pose an internal threat. To mitigate this threat, we conducted an exploratory experiment. Specifically, we select perturbed samples with $pd$=1 that LLMs are able to successfully repair as two new benchmark datasets, namely Defects4J$_{free-leakage}$ and QuixBugs$_{free-leakage}$ to redo the experiments in the Section \ref{5B}. The rationale behind this setting is that even if LLMs may have encountered the original samples in Defects4J$_{base}$ or QuixBugs$_{base}$ during pre-training, perturbed samples are much less likely to be seen by LLMs, allowing us to assess the influence of data leakage on this work. Results on Defects4J$_{free-leakage}$ and QuixBugs$_{free-leakage}$ are listed in Table \ref{leakage}, where the $R$-$score$s are relatively higher than those in Table \ref{tab:rotllm}. A potential explanation is if samples that LLMs have never seen can be repaired by LLMs, it means that LLMs have a better and deeper understanding of them, so the perturbation of such samples has little impact on the performance of LLMs. Besides, Table \ref{leakage} indicates that R-scores for LLMs on the new, leakage-free datasets show varying degrees of decline, which is similar in trend to results in Table \ref{tab:rotllm}, demonstrating that our conclusion obtained in this work is convincing and meaningful, and the threat is minimized.

\begin{table}[htbp]
\vspace{-1.5em}
\setlength{\abovecaptionskip}{0cm}
\caption{Data leakage testing}
\centering
\renewcommand{\arraystretch}{1.25}
\resizebox{\linewidth}{!}{
% \begin{threeparttable}
\begin{tabular}{lcccccc}
\hline
\multicolumn{1}{c}{\multirow{2}{*}{\textbf{Model}}} & \multicolumn{3}{c}{Defects4J$_{free-leakage}$}                                                                                    & \multicolumn{3}{c}{QuixBugs$_{free-leakage}$}                                                                                     \\ \cline{2-7} 
\multicolumn{1}{c}{}                                & \multicolumn{1}{l}{\textbf{Invalid}} & \multicolumn{1}{l}{\textbf{Valid}} & \multicolumn{1}{l}{\textbf{R-score}} & \multicolumn{1}{l}{\textbf{Invalid}} & \multicolumn{1}{l}{\textbf{Valid}} & \multicolumn{1}{l}{\textbf{R-score}} \\ \hline
Mistral Large                                       & 291                                  & 525                                & 0.643                                & 102                                  & 1848                               & 0.948                                \\
LLaMA3-70B                                          & 187                                  & 344                                & 0.648                                & 377                                  & 1381                               & 0.786                                \\
LLaMA3-8B                                           & 338                                  & 316                                & 0.483                                & 521                                  & 1278                               & 0.710                                \\
CodeGemma-7B                                        & 172                                  & 206                                & 0.545                                & 596                                  & 411                                & 0.408                                \\ \hline
\textbf{Avg. R\text{-}score}       & \multicolumn{3}{c}{0.585}                                                                                        & \multicolumn{3}{c}{0.755}                                                                                        \\ \hline
\end{tabular}}
\label{leakage}
\vspace{-0.5em}
\end{table}

\noindent\textbf{External Validity.}
% The selection of samples and LLMs may affect the accuracy and reliability of experimental results. 
The selection of base samples and LLMs may affect the conclusion. However, we follow a widely used taxonomy \cite{sobreira2018dissection} on buggy codes and try to make the selected samples cover all categories as much as possible to alleviate the former threat.  
% the sample selection method adopted in this study follows the previous practice of empirical research \cite{yuan2023no}, \cite{wang2024metmap}, ensuring rationality. 
Additionally, among the LLMs we selected, there are both open-source LLMs (LLaMA3, CodeGemma-7B) and closed-source LLMs (Mistral Large), and we also chose LLMs with different parameter sizes (LLaMA3-8B/70B). Besides, LLMs of diverse families are also included as above. Therefore, our evaluation of LLMs is relatively representative and can mitigate this threat. In the future, we can involve more LLMs in experiments. %Second, the effectiveness of JCRBooster is likely influenced by specific coding styles and programming practices prevalent in different codebases.    While our findings suggest that JCRBooster can improve code readability, the extent of this improvement may vary.    However, the coding styles and practices considered in our study are common in many software development environments, suggesting that the observed improvements in readability are broadly applicable.    Future research can explore the effectiveness of these models in a wider range of coding environments to further validate our conclusions.

\noindent\textbf{Construct Validity.}
This work adopts test suites offered in each dataset to assess the correctness of repaired programs without considering literal consistency to the ground truths, which may leave a threat to construct validity. Nonetheless, since perturbations may occur at the location of the original bug or affect their original fixing logic by accident, the correct repair result might not match the ground truths. To avoid these unexpected biases, we do not adopt a literal matching during the evaluation. Additionally, the current assessment has already revealed the robustness deficiencies of LAPR. Thus, we believe this threat is minimal.

\section{Conclusion}

%This study aimed to investigate the impact of code perturbations on the robustness of various defect repair models and to propose methods for improving model robustness through readability improvements.                    The objectives were to assess the significance of perturbations in real-world scenarios, analyze the robustness of models against different levels and types of perturbations, and develop strategies to mitigate the negative effects of perturbations on model performance.

This paper proposes the first comprehensive testing framework, MT-LAPR, to measure the robustness of LAPR techniques. Through a pilot study, we summarize nine widely accepted Metamorphic Relations (MRs) for MT-LAPR, and our evaluation shows that the test cases generated by the MRs we proposed can effectively identify the robustness deficiencies of LLM in automated program repair. Besides, we find there is a significant negative correlation between the performance of LAPR and perturbation distance, as posing more MR-based perturbations on buggy codes damages their readability. Motivated by this, we use test cases generated by MT-LAPR as samples to train a CodeT5-based code editing model aiming at improving buggy code readability and showing promising results in robustness improvements for LLMs. We believe that this work is the crucial first step toward systematic testing of LAPR techniques. For future work, we will continue developing metamorphic relations in MT-LAPR and extend it to various settings.

\bibliographystyle{IEEEtran}
\bibliography{sample}

\end{document}